\documentclass[aps,prb,amssymb,12pt,floatfix]{revtex4}
\usepackage{subfigure}
\usepackage{graphicx}
\usepackage{rotating}
\usepackage{amsmath, amsthm}

\begin{document}
\title{Kinetics of Interior Loop Formation in Semiflexible Chains}

\author{Changbong Hyeon and D. Thirumalai}
\affiliation{Biophysics Physics Program,
Institute for Physical Science and Technology\\
University of Maryland, College Park, MD 20742\\
}
\baselineskip = 22pt
\begin{abstract}
Loop formation between monomers in the interior of semiflexible chains 
describes elementary events in biomolecular folding and DNA bending. We calculate
analytically the interior distance distribution function for semiflexible 
chains using a mean-field approach.  Using the potential of mean force derived
from the distance distribution function we present a simple expression 
for the kinetics of interior looping by adopting Kramers theory. For the parameters, 
that are appropriate for DNA, the theoretical predictions in comparison to the case are in
excellent agreement with explicit Brownian dynamics simulations of worm-like
chain (WLC) model.  
The interior looping times ($\tau_{IC}$) can be greatly altered in cases when the stiffness of the loop differs 
from that of the dangling ends. 
If the dangling end is stiffer than the loop then $\tau_{IC}$ increases for the case of 
the WLC with uniform persistence length. 
In contrast, attachment of flexible dangling ends enhances rate of interior loop formation. 
The theory also shows that if the monomers are charged 
and interact via screened Coulomb potential then
both the cyclization ($\tau_c$) and interior looping ($\tau_{IC}$)
times greatly increase at low ionic concentration. Because both
$\tau_c$ and $\tau_{IC}$ are determined essentially by the effective 
persistence length ($l_p^{(R)}$)  we computed $l_p^{(R)}$ by varying the
range of the repulsive interaction between the monomers.  For short range
interactions $l_p^{(R)}$ nearly coincides with the bare persistence length  which
is determined largely by the backbone chain connectivity. This finding
rationalizes the efficacy of describing a number of experimental
observations (response of biopolymers to force and cyclization kinetics) in
biomolecules using WLC model with an effective persistence length. 
\end{abstract}
\maketitle

\section{Introduction}

The kinetics of formation of contact between the ends of a polymer chain has a rich history.\cite{Florybook2,WinnikBook} 
Both experiments,\cite{Florybook2,WinnikBook} theory,\cite{FixmanJCP73,FixmanJCP74,PastorJCP96,DoiCP75,SzaboJCP80,PericoJCP81,ThirumalaiJPCB99} and simulations \cite{PericoPPS84,ReyMAC91,VologodskiiMacro97} 
have been used to address the elementary event of the 
dynamics of end-to-end contact formation (or cyclization kinetics) (Fig.\ref{looping.fig}-A). 
Contact formation between two reactive groups separated by a certain distance along the chain is a 
basic intramolecular rate process in a polymer.
Recently, there has been renewed interest in understanding 
the looping dynamics that has been studied both theoretically 
\cite{FixmanJCP74,DoiCP75,SzaboJCP80,PericoJCP81,PastorJCP96} 
and experimentally \cite{BaldwinPNAS81,EatonJPC97,HofrichterJPCB02,VologodskiiMacro97,VologodskiiPNAS05} 
because of its fundamental importance in a number of biological processes. 
The hairpin loop formation is the elementary step in RNA folding,\cite{TinocoJMB99} structure formation in ssDNA,\cite{LibchaberPNAS98,AnsariBJ01}
and protein folding.\cite{EatonJPC97,KlimovPNAS00,KiefhaberPNAS99,GrayPNAS03,KiefhaberJMB05}
Cyclization in DNA has recently drawn renewed attention not only 
because of its importance in gene expression \cite{LangowskiBJ98,HalfordARBBS04} but also it provides a way to assess DNA's flexibility.
The promise of using single molecule technique to probe the real time dynamics of polymer chains
has also spurred theories and simulations of cyclization kinetics. 
Using loop formation times 
between residues that are in the interior as the most elementary event in 
protein folding, it has been argued, using experimental data and theoretical expression for probability for 
loop formation in stiff chains,
that the speed limit for folding is on the order of a 1 $\mu s$.\cite{EatonPNAS96}
These examples illustrate the need to understand quantitatively the elementary event of contact formation between segments of a polymer chain.

Even without taking hydrodynamic interactions into account theoretical treatment of cyclization kinetics in polymer chains is difficult 
because several relaxation times and length and energy  scales are interwined. 
At the minimum the variation of time scale for cyclization ($\tau_c$) with polymer length is dependent on polymer relaxation time ($\tau_R$). 
In biopolymers additional considerations due to chain stiffness and heterogeneity of interactions between monomer (amino acid residue or nucleotides) must 
be also taken into account. Majority of the cyclization kinetics studies on synthetic polymers \cite{WinnikBook}
have considered examples in which the contour length ($L$) of the polymer is much 
greater than its persistence length ($\l_p$). 
In contrast, loop formation dynamics in biopolymers have focused on systems in which $L/l_p$ is relatively small. 
In disordered polypeptide chains $L/l_p$ can be as small as $3$,\cite{HofrichterJPCB02,HudginsJACS02} while 
in DNA $L/l_p<1$.\cite{WidomMC04,VologodskiiPNAS05}
Thus, it is important to develop theoretical tools for the difficult problem of loop formation 
dynamics for arbitrary $L$ and $l_p$. 
Despite the inherent complexities in treating loop formation in biopolymers it has been found that the use of 
polymer-based approach is reasonable in analyzing experimental data 
on cyclization kinetics in proteins \cite{ThirumalaiJPCB99,HofrichterJPCB02} and DNA.\cite{VologodskiiPNAS05}

In this paper we are primarily concerned with the looping dynamics  between interior 
segments of a semiflexible chain. 
While a lot of theoretical and experimental works (mentioned above) have been done on the 
end-to-end looping (Fig.\ref{looping.fig}-(a)), only a few studies have been reported 
on the contact formation between monomers in the interior of a chain (interior looping) (Fig.\ref{looping.fig}-(b)).\cite{WinnikMacro97,PericoMacro90,FriedmanJPII91,CamachoPROTEINS95,EatonPNAS93}
There are a few reasons to consider kinetics of interior looping. 
(1) The biological events such as hairpin formation and DNA looping often involve contact formation between monomers that are not at the ends of the chain. 
For example, it is thought that the initiation of nucleation in protein folding occurs 
at residues that are near the loop regions.\cite{GuoBP95}
The residues that connect these loops are in the interior of the polypeptide chain. 
Similar processes are also relevant in RNA folding.\cite{HyeonBC05}
(2) It is known that for flexible chains with excluded volume interactions (polymer in a good solvent) the probability of loop formation is strongly dependent on the location of the two segments. 
For large loop length ($S$) the loop formation probability, 
$P(S)$, in three dimensions for chain ends $\sim S^{\theta_1}$ where $\theta_1\approx 1.9$ while 
$P(S)\sim S^{\theta_2}$ with $\theta_2\sim 2.1$ for monomer in the interior.\cite{desCloizeauxPRA74}
Although the values of $\theta_1$ and $\theta_2$ are similar it could lead to measurable differences in loop formation times.\cite{CamachoPROTEINS95} 
 
The rest of the paper is organized as follows. 
In section II we present the physical considerations that give rise to the well-established results for $\tau_c$ for flexible chains. 
The extension of the arguments for flexible chains to semiflexible polymers suggests that the local 
equilibrium approximation can be profitably used to analyze both 
cyclization kinetics and interior looping dynamics. 
The basic theory for the equilibrium distance distribution between two interior segments $s_1$ and $s_2$ 
(Fig.\ref{looping.fig}-(b)) is presented in section III. 
Using the equilibrium distribution function and adopting Kramers theory and following the suggestion by Jun \emph{et. al.},\cite{HaELett03}
we obtain an analytical expression for time scale $\tau_{IC}$ for interior contact formation in section IV. 
Explicit results of simulations of worm-like chain (WLC), which validate the theory, are presented in section V. 
In section VI we consider the kinetics of interior loop formation in WLC in which the stiffness of the loop is different from that of the dangling ends. 
Section VII describes the consequences of screened Coulomb interaction between monomer segments 
on cyclization kinetics and interior looping dynamics. 
Because the results in section VIII are expressed in terms of a renormalized persistence length ($l_p^{(R)}$) of WLC 
we present simulation results for $l_p^{(R)}$ variation for a number of potentials that describe interactions between monomers in section VII. 
The conclusions of the article are summarized in section IX. 
\\

\section{Preliminary considerations}

The pioneering treatment of loop formation dynamics due to Wilemski and Fixman (WF) \cite{FixmanJCP74,FixmanJCP74II} has formed 
the basis for treating cyclization kinetics in flexible polymer chains.
Using a generalized diffusion equation for the probability density,
 $\frac{\partial P(\{\textbf{r}^N\},t)}{\partial t}=\mathcal{L}_{FP}P(\{\textbf{r}^N\},t)-k\mathcal{S}(\{\textbf{r}^N\})P(\{\textbf{r}^N\},t)$ 
($\mathcal{L}_{FP}$ is a generalized diffusion operator, $k\mathcal{S}$ is a sink term)
for a N-segment polymer, and local equilibrium approximation within the sink,  
WF expressed the cyclization time $\tau_c$ in terms of an integral involving a sink-sink correlation function. 
From the WF formalism and related studies it is known 
that even in the simplest cases (ideal chains or polymers with excluded volume interactions) 
the validity of the local equilibrium approximation depends on the interplay between $\tau_c$ and the chain relaxation time, $\tau_R$. 
If $\tau_c\gg\tau_R$ then the local equilibrium approximation is expected to hold because the polymer chain effectively explores the available volume \emph{before} the monomers at the end (reactive groups) form a contact. 
In this situation, $\tau_c$ can be computed by considering mutual diffusion of the chain ends in a potential of mean force 
($F(R_e)$). 
For ideal chains, $F(R_e)=-k_BT\log{P(R_e)}\sim3k_BTR_e^2/2\overline{R}^2$ where $R_e$ is the end-to-end distance, $\overline{R}\sim aN^{1/2}$ is the mean end-to-end distance, $a$ is the 
size of the monomer, $T$ is the temperature, and $k_B$ is the Boltzmann constant. 
By solving such an equation subject to the absorbing boundary condition, Szabo, Schulten, and Schulten (SSS) \cite{SzaboJCP80} showed 
that $\tau_{SSS}=\tau_oN^{3/2}$. 
Simulations \cite{PastorJCP96} and theory \cite{DoiCP75} show that if the capture radius for contact formation is non-zero, and is on the order of a 
monomer size 
then $\tau_c\sim\frac{\langle R^2\rangle}{D_c}\sim \tau_1N^{2\nu+1}$ where $\nu=1/2$ for Rouse chains and $\nu\approx 3/5$ for polymers with 
excluded volume, and $D_c$ is a mutual diffusion coefficient. 
The use of these theories to analyze the dependence of $\tau_c$ on $N$ in polypeptides shows that the physics of cyclization kinetics is reasonably well described by diffusion in a potential of mean force $F(R_e)$ 
which only requires accurate calculation of $P(R_e)$ the end-to-end distribution function.\cite{HofrichterJPCB02,HaELett03} 
For describing interior looping times $\tau_{IC}$ for contact between two interior monomers $s_1$ and $s_2$ we need to compute $P(R_{12},|s_1-s_2|)$ 
where $R_{12}$ is the distance between $s_1$ and $s_2$. 
With $P(R_{12},|s_2-s_1|)$ in hand 
$\tau_c$ can be computed by solving a suitable diffusion equation. 

Because the use of $F(R_e)$ in computing $\tau_c$ and $\tau_{IC}$ is intimately related to chain relaxation times it is useful to survey the conditions which satisfy the local equilibrium approximation. 
By comparing the conformational space explored by the chain ends compared to the available volume prior to cyclization \cite{deGennesJCP82} the validity of the local equilibrium 
approximation in flexible chains can be expressed in terms of an exponent $\theta=\frac{d+g}{z}$.\cite{PericoJCP81}
Here $d$ is the spatial dimension, the correlation hole exponent (des Cloizeaux exponent) \cite{desCloizeauxJP80} 
$g$ describes the probability of the chain 
ends coming close together, and $z$ is the dynamical scaling exponent 
($\tau_R\sim \overline{R}^z$).
If $\theta>1$ the local equilibrium approximation is expected to hold and $\tau_c$ is 
determined essentially by the equilibrium $P(R_e)$ as $R_e\rightarrow a$ the capture radius. 
Using the scaling form of $P(R_e)$ for small $R_e$ $P(R_e)\sim\frac{1}{\overline{R}^d}\left(\frac{R_e}{\overline{R}}\right)^g$ and $\overline{R}\sim N^{\nu}$ ($\nu$ is the Flory exponent) 
we find $\tau_c\sim N^{\nu(d+g)}$. 
For Gaussian chains $\nu=1/2$ and $g=0$ and hence $\tau_{SSS}\sim\tau_c\sim N^{3/2}$. 
This result was obtained fifty years ago by Jacobsen and Stockmayer.\cite{StockmayerJCP50}
However, in the free-draining case ($z=4$, $g=0$, $\nu=1/2$, $d=3$), 
$\theta<1$ and hence the condition $\tau_c\gg\tau_R$ is not satisfied. 
In this case $\tau_c\sim\tau_R\sim N^{z\nu}\sim N^2$. 
Thus, for ideal Gaussian chains 
it is likely that $\tau_{SSS}<\tau_c<\tau_{WF}$.\cite{PortmanJCP03}
Indeed, recent simulations show that if the number of statistical segments is large ($\gtrsim 20$) then for ideal 
chains $\tau_c\sim N^2$ which signals the breakdown of the condition $\tau_c\gg\tau_R$. 
Experiments on cyclization of polypeptide chains show that $\tau_c\sim N^{3/2}a$ is obeyed for $N$ in the range $10<N<20$ 
(see Fig.(5) in Ref.\cite{HofrichterJPCB02}). 
Deviations from ideal chain results are 
found for $N<10$, either due to chain stiffness \cite{HofrichterJPCB02} or sequence variations.\cite{NauRCI05}
For polymer chains in good solvents with hydrodynamic interactions ($d=3$, $g=5/18$, $z=3$, and $\nu=3/5$), 
$\theta=59/54>1$. Thus, in real chains the local equilibrium approximation may be accurate. 

For stiff chains bending rigidity severely restricts the allowed conformations especially when the 
contour length ($L$) is on the order of the persistence length ($l_p$). 
Because of high bending rigidity the available volume is restricted by thermal fluctuations. 
Clearly in this situation, the chain is close to equilibrium. 
This may be the case for short DNA segments. 
In effect these chains satisfy the $\tau_c>\tau_R$ condition which enables 
us to calculate $\tau_c$ or $\tau_{IC}$ by solving 
an appropriate one dimensional diffusion equation (see below) in a suitable potential of mean force.

\emph{Effect of chain Stiffness :}
Many biopolymers are intrinsically stiff and are better described by worm-like chain (WLC) models. 
The persistence length, which is a measure of stiffness, varies considerably. 
It ranges from $(3-7)$ \AA\ (proteins),\cite{HofrichterJPCB02} $(10-25)$ \AA\ (ss-DNA \cite{AnsariBJ01,WeissPNAS05} and RNA,\cite{CaliskanPRL05,HyeonBJ06}) 
50 $nm$ for ds-DNA. 
Typically, loops of only a few persistence length form, 
which underscores the importance of chain stiffness. 
In order to correctly estimate the loop closure time, consideration of 
the stiffness in the loop closure dynamics is necessary
unless the polymer looping takes place between the reactive groups that are well 
separated and the chain length $L$ is long. 
If $L\gg l_p$ (persistence length), the looping dynamics will follow the scaling law for flexible chains. 
However, at short length scales loop dynamics can be dominated by chain stiffness.\cite{VologodskiiPNAS05}
If the chain is stiff then WLC conformations are limited to those allowed by thermal fluctuations. 
In this situation, the time for exploring the chain conformations is expected to be less 
than $\tau_c$. 
Thus, we expect local equilibrium to be a better approximation for WLC 
than for long flexible chains.

Recently, Dua \emph{et. al.}\cite{CherayilJCP02} have studied the effect of stiffness on the polymer dynamics 
based on Wilemski-Fixman formalism and showed, that for 
free-draining semiflexible chain without excluded volume 
$\tau_c\sim N^{2.2\sim 2.4}$ at moderate values of stiffness. 
However, the procedure used to obtain this result is not complete, as recognized by the authors, 
because they use a Gaussian propagator
$\mathcal{G}(\textbf{r},t|\textbf{r}',0)=\left(\frac{3}{2\pi\langle\textbf{r}^2\rangle(1-\phi(t))}\right)^{3/2}\exp{\left(-\frac{3(\textbf{r}-\phi(t)\textbf{r}')^2}{2\langle\textbf{r}^2\rangle(1-\phi(t))}\right)}$ which is not valid for WLC. 
The end-to-end distance distribution becomes a Gaussian at equilibrium,  
$\lim_{t\rightarrow\infty}\mathcal{G}(\textbf{r},t|\textbf{r}',0)=P_{eq}(\textbf{r})=\left(\frac{3}{2\pi\langle\textbf{r}^2\rangle}\right)^{3/2}\exp{\left(-\frac{3\textbf{r}^2}{2\langle\textbf{r}^2\rangle}\right)}$, 
which is incorrect for semiflexible chain especially when $l_p\sim L$ (see Fig.\ref{Dfunc.fig} and Refs.\cite{BhattaThirumCondmat97, FreyPRL96}). 

As an alternative method 
we include the effect of chain stiffness assuming that local equilibrium approximation is valid. 
This is tantamount to assuming that $\tau_c>\tau_R$ which, for reasons given above, may be an excellent 
approximation 
for WLC.\cite{HaELett03}
In this case we can compute $\tau_c$ by solving the diffusion equation in a one dimensional potential 
$F(R_e)=-k_BT\ln{P(R_e)}$ where $P(R_e)$ is the probability of end-to-end distance distribution for WLC. 
For the problem of interest, namely, the computation of $\tau_{IC}$, 
we generalize the approach of Jun \emph{et. al.} \cite{HaELett03} 
who used Kramers theory in the effective potential $F(R_e)$ to obtain $\tau_c$. 
In general, the time for cyclization can be calculated using 
\begin{equation}
\tau_c=\int^r_{a} dye^{\beta F(y)}\frac{1}{D}\int_y^Ldze^{-\beta F(z)}
\label{eqn:mfpt}
\end{equation}
where $a$ is the capture (contact) radius of the two reactive groups. 
We show that Eq.(\ref{eqn:mfpt}) provides accurate estimates of $\tau_c$, thus suggesting the local equilibrium approximation is guaranteed.

Here, we address the following specific questions:
What is the loop formation time between the interior segments in a semiflexible chain?
Does the dangling ends (Fig.\ref{looping.fig}-(b)) affect the dynamics of loop formation? 
How does the effect of interaction between monomer segments 
(e.g. excluded volume, electrostatic interaction) affect loop closure kinetics in WLC models? \\

\section{Distance distribution function between two interior points}

A key ingredient in the calculation of the potential of mean force is 
appropriate distribution function between the two monomers that form a contact. 
In Refs.\cite{BhattaThirumCondmat97,HaBook} the equilibrium end-to-end ($R_e$) 
radial distribution function 
of a semiflexible chain $P(R_e)$ was obtained in terms of the persistence length ($l_p$) 
and the contour length ($L$).
Despite the mean field approximation employed in Refs.\cite{BhattaThirumCondmat97,HaBook} the distribution function 
$P(R_e)$ is in very good agreement with simulations.\cite{FreyPRL96}
The simplicity of the final expression has served as a basis for analyzing a number of experiments on proteins,\cite{EatonPNAS05}
RNA \cite{CaliskanPRL05} and DNA.\cite{TJHaBJ04}
In this section, we use the same procedure to calculate the distribution function $P(R_{12};l_p,s_1,s_2,L)$ where $0<s_1,s_2<L$, and $R_{12}$ is the spatial distance between $s_1$ and $s_2$.

For the semiflexible chain in equilibrium we write  
the distribution function of the distance $R_{12}$ between $s_1$ and $s_2$ (Fig.\ref{looping.fig}) along the chain contour as 
\begin{eqnarray}
  G(\textbf{R}_{12};s_1,s_2)&=&\langle\delta(\textbf{R}_{12}-\int^{s_2}_{s_1}\textbf{u}(s)ds)\rangle_{MF}\nonumber\\
  &=&\frac{\int D[\textbf{u}(s)]\delta(\textbf{R}_{12}-\int^{s_2}_{s_1}\textbf{u}(s)ds)\Psi_{MF}[\textbf{u}(s)]}{\int D[\textbf{u}(s)]\Psi_{MF}[\textbf{u}(s)]}
\label{eqn:pathintegral}
\end{eqnarray}
where $\textbf{u}(s)$ is a unit tangent vector at position $s$.
The exact weight for the semiflexible chain is 
$\Psi[\textbf{u}(s)]\varpropto\exp{[-\frac{l_p}{2}\int^L_0ds\left(\frac{\partial\textbf{u}}{\partial s}\right)^2}]\prod\delta(\textbf{u}^2(s)-1)$.
The nonlinearity, that arises due to the restriction $\textbf{u}^2(s)=1$, makes the computation of the path integral in 
Eq.(\ref{eqn:pathintegral}) difficult. To circumvent the problem we replace $\Psi[\textbf{u}(s)]$ by the mean field weight $\Psi_{MF}[\textbf{u}(s)]$, \cite{LNN91}
\begin{equation}
  \Psi_{MF}[\textbf{u}(s)]\varpropto
  \exp{[-\frac{l_p}{2}\int^L_0\left(\frac{\partial\textbf{u}(s)}{\partial s}\right)^2ds-\lambda\int^L_0(\textbf{u}^2(s)-1)ds-\delta[(\textbf{u}^2_0-1)+(\textbf{u}^2_L-1)]]}. 
  \label{eqn:weight}
\end{equation}
The Lagrange multipliers $\lambda$ and $\delta$, which are used to enforce the constraint $\textbf{u}^2(s)=1$,\cite{HaJCP95}
will be determined using stationary phase approximation (see below). 
The path integral associated with the weight $\Psi_{MF}[\textbf{u}(s)]$ is equivalent to a kicked quantum mechanical harmonic oscillator with ``mass'' $l_p$ and 
angular frequency $\Omega\equiv\sqrt{2\lambda/l_p}$. 
Using the propagator for the harmonic oscillator
\begin{equation}
Z(\textbf{u}_s,\textbf{u}_0,s)=\left(\frac{\pi\sinh{(\Omega s)}}{\Omega_p}\right)^{-\frac{3}{2}}
  \exp{(-\Omega_p\frac{(\textbf{u}_s^2+\textbf{u}_0^2)\cosh{(\Omega s)}-2\textbf{u}_s\cdot\textbf{u}_0}{\sinh{(\Omega s)}})}
\label{eqn:Z}
\end{equation}
and defining $\Omega_p\equiv\frac{\Omega l_p}{2}$ the isotropic distribution function becomes 
\begin{eqnarray}
 G(R_{12},s_1,s_2)
&=&\mathcal{N}^{-1}e^{\lambda L+2\delta}\int\frac{d^3\textbf{k}}{(2\pi)^3}\int d\textbf{u}_0d\textbf{u}_{s_1}d\textbf{u}_{s_2}d\textbf{u}_Le^{-\delta\textbf{u}_0^2}Z(\textbf{u}_0,\textbf{u}_{s_1};s_1)\nonumber\\
&\times&e^{i\textbf{k}\cdot\textbf{R}_{12}-\frac{\textbf{k}^2}{4\lambda}|s_1-s_2|}Z(\textbf{u}_{s_1}+\frac{i\textbf{k}}{2\lambda},\textbf{u}_{s_2}+\frac{i\textbf{k}}{2\lambda};s_2-s_1)\nonumber\\
&\times&e^{-\delta\textbf{u}_L^2}Z(\textbf{u}_{s_2},\textbf{u}_L;L-s_2).
\end{eqnarray}
By writing the distribution function as $ G(R_{12},s_1,s_2)=\int^{i\infty}_{-i\infty}d\lambda\int^{i\infty}_{-i\infty}d\delta\exp{(-\mathcal{F}[\lambda,\delta])}$
it is clear that the major contribution to $G$ (in the thermodynamic limit $L\rightarrow\infty$)
comes from the saddle points of the free energy functional 
$\mathcal{F}[\lambda,\delta]$, i.e., 
$\frac{\partial\mathcal{F}}{\partial\lambda}=\frac{\partial\mathcal{F}}{\partial\delta}=0$. 
The functional $\mathcal{F}[\lambda,\delta]$  is (see Appendix A for details of the derivation)
\begin{eqnarray}
\mathcal{F}[\lambda,\delta]&=&-(L\lambda+2\delta)\nonumber\\
&+&\frac{3}{2}\ln{\left(\frac{\sinh{\Omega L}}{\Omega_p}(\delta^2+\Omega_p^2+2\delta\Omega_p\coth{\Omega L})\right)}-\frac{3}{2}\ln{\frac{\lambda^2}{Q(s_1,s_2;\lambda,\delta)}}\nonumber\\
&+&\frac{\lambda^2R^2_{12}}{Q(s_1,s_2;\lambda,\delta)}
\label{eqn:functional_result}
\end{eqnarray}

To obtain the optimal values of $\lambda$ and $\delta$ we first take the $L\rightarrow\infty$ limit and then solve stationarity conditions 
$\frac{\partial\mathcal{F}}{\partial\lambda}=\frac{\partial\mathcal{F}}{\partial\delta}=0$. 
Technically, the optimal value of $\delta$ and $\lambda$ should be calculated for a given $L$ and then it is 
proper to examine the $L\rightarrow \infty$ limit. 
The consequences of reversing the order of operation are discussed in Appendix B. Using the first procedure (taking $L\rightarrow\infty$ first) 
we obtain 
$Q(s_1,s_2;\lambda,\delta)\rightarrow|s_2-s_1|\lambda$ 
in the limit $L\gg s_2\gg s_1\gg1$, and thus $\mathcal{F}[\lambda,\delta]$ becomes 
\begin{eqnarray}
\mathcal{F}[\lambda,\delta]&\approx&-(L\lambda+2\delta)\nonumber\\
&+&\frac{3}{2}\ln[\frac{e^{\Omega L}}{\Omega_p}(\frac{\delta}{\Omega_p}+1)^2]+\frac{3}{2}\ln{\Omega_p^2}+\frac{3}{2}\ln{\frac{|s_2-s_1|}{\lambda}}+\frac{\textbf{R}^2_{12}\lambda}{|s_2-s_1|}\nonumber\\
&=&L\left(\frac{3}{2}\Omega-\lambda(1-\frac{|s_2-s_1|}{L}\frac{\textbf{R}_{12}^2}{|s_2-s_1|^2})\right)\nonumber\\
&+&\frac{3}{2}\ln{[\frac{\Omega_p}{\lambda}(\frac{\delta}{\Omega_p}+1)^2]}+\frac{3}{2}\ln{|s_2-s_1|}-2\delta
\label{eqn:Functional}
\end{eqnarray}
where we have omitted numerical constants. 
The major contribution to the integral over $\lambda$ and $\delta$ comes from the sets of $\lambda$ and $\delta$ which pass the saddle point of a stationary phase contour on the 
$\mathbf{Re}\{\mathcal{F}\}$ plane.
Since the term linear in $L$ dominates the logarithmic term in
$l_p$ even when $L/l_p\sim\mathcal{O}(1)$, the stationary condition for 
$\lambda$ can be found by taking the derivative with respect to $\lambda$ by considering only 
the leading term in $L$ (cf. see Appendix B for details). 
The stationarity condition leads to 
\begin{equation}
\Omega_p=\sqrt{\frac{\lambda l_p}{2}}=\frac{3}{4}\frac{1}{1-\frac{|s_2-s_1|}{L}\textbf{r}^2}
\label{eqn:stationarycondition}
\end{equation}
where $\textbf{r}=\frac{\textbf{R}_{12}}{|s_2-s_1|}$ with $0<r<1$.
Similarly, the condition for $\delta$ can be obtained as 
\begin{equation}
\delta=\frac{3}{2}-\Omega_p.
\end{equation}
Determination of the parameters $\lambda$ and $\delta$ by the stationary phase approximation amounts to replacing the local constraint $\textbf{u}^2(s)=1$ by a 
global constraint $\langle\textbf{u}^2(s)\rangle=1$.\cite{HaJCP95}
Finally, the stationary values of $\lambda$ and $\delta$ in the large $L$ limit 
give the interior distance distribution function:
\begin{equation}
G(R_{12},s_2-s_1)=\frac{N}{(1-\frac{|s_2-s_1|}{L}r^2)^{9/2}}\exp{(-\frac{9|s_2-s_1|}{8l_p(\frac{|s_2-s_1|}{L})(1-\frac{|s_2-s_1|}{L}r^2)})}.
\end{equation}

The mean-field approximation allows us to obtain a simple expression for the internal segment distance 
distribution function. 
The previously computed $P(R_e)$ \cite{BhattaThirumCondmat97} can be retrieved by setting $|s_2-s_1|=L$. 
The radial probability density, for the interior segments, in three dimensions, for semiflexible chains is 
\begin{equation}
P(r;s_2-s_1,t)=4\pi C\frac{r^2}{(1-\frac{|s_2-s_1|}{L}r^2)^{9/2}}\exp{(-\frac{3t}{4(\frac{|s_2-s_1|}{L})(1-\frac{|s_2-s_1|}{L}r^2)})}
\label{eqn:PofR}
\end{equation}
where $r=R_{12}/|s_2-s_1|\equiv R/|s_2-s_1|$ and $t=|s_2-s_1|/l_0$ with $l_0=\frac{2}{3}l_p$.
The normalization constant $C$ is determined using $\int^1_0P(r,s;t)dr=1$.
The integral is evaluated by the substitution $\sqrt{\frac{|s_2-s_1|}{L}}r=\frac{x}{\sqrt{1+x^2}}$ to yield
\begin{eqnarray}
C&=&\frac{1}{4\pi}(\frac{|s_2-s_1|}{L})^{3/2}\left(\int_0^{x_0}dxx^2(1+x^2)e^{-\alpha(1+x^2)}\right)^{-1}\nonumber\\
&=&\frac{4}{\pi\alpha^{-7/2}}(\frac{|s_2-s_1|}{L})^{3/2}
[-2\sqrt{\alpha}x_0e^{-\alpha(1+x_0^2)}(15+2\alpha(6\
+5x_0^2+2\alpha(1+x_0^2)^2))\nonumber\\
&+&\alpha^2e^{-\alpha}\sqrt{\pi}\mathrm{erf}{[\sqrt{\alpha}x_0]}(1+3\alpha^{-1}+\frac{15}{4}\alpha^{-2})]^{-1}
\end{eqnarray}
where $\alpha=\frac{3t}{4\frac{|s_2-s_1|}{L}}$, $x_0=\sqrt{\frac{|s_2-s_1|}{L-|s_2-s_1|}}$, and $\mathrm{erf}(x)$ is the error function. 
The peak in the distribution function is at
\begin{equation}
r_{max}=\sqrt{\frac{\eta+\sqrt{\eta^2+14}}{7|s_2-s_1|/L}}
\label{eqn:rmax}
\end{equation}
where $\eta=\frac{5}{2}-\frac{3t}{4\frac{|s_2-s_1|}{L}}$.
For $|s_2-s_1|=L$, $r_{max}\rightarrow 0$ as $t\rightarrow\infty$ and  $r_{max}\rightarrow 1$ as $t\rightarrow 0$.

In Fig.\ref{Dfunc.fig} we compare the distribution functions $P(R_e)$ and $P(R_{12},s_2-s_1)$.
When $|s_1-s_2|/L=1$ Eq.(\ref{eqn:PofR}) gives the end-to-end distribution for semiflexible chains. 
By adjusting the value of $t$ (or equivalently $l_0$) we can go from flexible to intrinsically stiff chains. 
As the chain gets stiff there is a dramatic difference between the $P(r;|s_1-s_2|,t)$ and $P(r;|s_1-s_2|=L,t)$ (see Fig.\ref{Dfunc.fig}-(b)).
Contact formation between interior segments are much less probable than cyclization process (compare the green and red curves with the black in Fig.\ref{Dfunc.fig}-(b)). 
Physically, this is because stiffness on shorter length scales ($|s_1-s_2|/L<1$) is more severe than when 
$|s_1-s_2|/L\sim\mathcal{O}(1)$. 
However, when the chain is flexible (large $t$) the difference between the probability of contact between the interior segments and cyclization is small (Fig.\ref{Dfunc.fig}-(a)). 
In the limit of large $t(\propto L/l_p)$ the Hamiltonian in Eq.(\ref{eqn:weight}) describes a Gaussian 
chain for which the distance distribution between interior points remains a Gaussian. 
However, if excluded volume interactions are taken into account there can be substantial difference between 
$P(x,|s_1-s_2|,t)$ and end-to-end segment distribution even when $t$ is moderately large. \\

\section{Interior loop closure time using Kramers theory}

Having obtained the effective potential between interior segments of semiflexible chain
we can evaluate Eq.(\ref{eqn:mfpt}) using $F(R)=-k_BT\log{P(R)}$ with $P(R)$ given by Eq.(\ref{eqn:PofR}) with $r=R/L$.
For clarity we have suppressed the dependence of $P(R)$ on $|s_2-s_1|$. 
The expression for the mean first passage time (Eq.(\ref{eqn:mfpt})) can be approximated by expanding the effective potential $F(R)$ 
at the barrier top and at the the bottom as 
$F(R)\approx F(R_t)-\frac{1}{2}\nabla_R^2F(R)|_{R=R_t}(R-R_t)^2+\cdots$ 
and $F(R)\approx F(R_b)+\frac{1}{2}\nabla_{R}^2F(R)|_{R=R_b}(R-R_b)^2+\cdots$, respectively (Fig.\ref{Dfunc.fig}). 
Evaluating the resulting Gaussian integrals yields the Kramers equation 
\begin{equation}
\tau_c\sim\tau_{Kr}=\frac{\pi k_BT}{D\sqrt{\nabla^2_RF(R)|_{R=R_b}}\sqrt{\nabla^2_RF(R)|_{R=R_t}}}\exp{(\Delta F^{\ddagger}/k_BT)}. 
\end{equation}
When evaluating the Gaussian integral at the barrier top with $R=R_t$ (Fig.\ref{Dfunc.fig}), 
we assume that only the integral beyond $R>R_t$ contributes to the result. 
In the overdamped limit the mean first passage time, 
which is roughly the inverse of the reaction rate, is determined by the barrier height ($\Delta F^{\ddagger}=F(R_t)-F(R_b)$), and 
the curvatures of the bound state, the curvature at the barrier top, 
and the friction coefficient, 
that depends on $D(=2D_0)$ where $D_0$ is the monomer diffusion coefficient.
The curvatures of the potential at the bottom ($R_{\alpha}=R_b$) or at the top ($R_{\alpha}=R_t$) (Fig.\ref{Dfunc.fig} right panel) is obtained using 
$\sqrt{\nabla^2_RF(R)|_{R=R_{\alpha}}}=\sqrt{k_BT\left(\frac{6}{R^2_{\alpha}}-\frac{G''(R_{\alpha},S,L)}{G(R_{\alpha},S,L)}\right)}$
by imposing the condition $\nabla_RF(R)|_{R=R_{\alpha}}=0$, i.e., $\left(\frac{2}{R}+\frac{\partial_RG(R,S,L)}{G(R,S,L)}\right)|_{R=R_{\alpha}}=0$.
There is an uncertainty in the evaluation $\sqrt{\nabla^2_RF(R)|_{R=R_t}}$ 
because $F(R)$ does not really form a barrier at $R=R_t$. 
Thus, we assume that the curvature at the barrier top is $\sim 1/R_t$ using dimensional analysis.  
We express lengths in terms of the persistence length 
$l_p^{(0)}=l_0$. 
Setting $\frac{|s_1-s_2|}{l_p^{(0)}}\equiv \textit{s}$, 
$\frac{L}{l_p^{(0)}}\equiv \textit{l}$, and $\frac{R}{l_p^{(0)}}\equiv\textit{x}$ the radial probability density is  
\begin{equation}
P(x,s,l)=\frac{4\pi C(s,l)(\frac{x^2}{s^3})}{(1-\frac{x^2}{ls})^{9/2}}\exp{[-\frac{3s}{4(s/l)(1-\frac{x^2}{ls})}]}.
\end{equation}
with $\int^s_0dxP(x,s,l)=1$.
When the dimensionless contact radius $x_t\equiv\alpha(=a/l_p^{(0)})\ll 1$, the exponential factor can 
be approximated as $\exp{(-\Delta F^{\ddagger}/k_BT)}=P(\alpha,s,l)/P(x_b,s,l)\simeq\frac{\alpha^2G(0,s,l)}{x_b^2G(x_b,s,l)}$.

The function $P(r)$ (Eq.(\ref{eqn:PofR})) is not appropriate for estimating the contact probability of semiflexible chains even though the 
overall shape of the mean field distribution function is in excellent agreement with the simulations and experiment. 
The contact probability for DNA is well studied by Shimada and Yamakawa,\cite{YamakawaMacro84}
thus we use their result for $G(0,l)$. 
If $x=0$ and $s<10$ then the Shimada-Yamakawa equation
gives a reliable estimate of the  
\emph{looping probability} ($G(0,s,l)=G_0(s)$). 
\begin{equation}
G_0(s)=\frac{896.32}{s^5}\exp{(-14.054/s+0.246s)}
\label{eqn:G0}
\end{equation} 
At a large $s$($>10$) value an interpolation formula $G_0(s)\sim s^{-3/2}$ due to 
Ringrose \emph{et. al.}\cite{RingroseEMBO} can be used. 
Note that as the chain gets stiffer ($l$ decrease) $G_0(s)$ decreases substantially indicating a great reduction in the loop formation probability for intrinsically stiff chains. 
It should be stressed that Eq.(\ref{eqn:G0}) has been obtained only for cyclization process and does not take into 
account the effect of dangling ends (Fig.\ref{looping.fig}-(b)). 
The contact probability ($r\rightarrow 0$) between interior segments should be different from the one for the end-to-end contact. In other words $G(r\rightarrow 0,s,l)$ should depend on $s/l$. 
Unfortunately, we do not know of analytical results for $G(0,s,l)$. 
We simply use $G_0(s)$ for $G(0,s,l)$ and resort to the values of $x_b$ and $G(x_b,s,l)$ to account for end effects. 
We validate the approximation that 
$G(0,s,l)$ does not depend on $l$ ($G_0(s)=G(0,s,l)$) 
explicitly using simulations (see below).

With these approximations the loop formation time in the presence of dangling ends (Fig.\ref{looping.fig}-(b)) is 
\begin{equation}
\tau_c(s,l)\simeq\{\frac{\pi}{\xi} x_b^2G(x_b,s,l)/(\frac{6}{x_b^2}-\frac{G''(x_b,s,l)}{G(x_b,s,l)})^{1/2}\}\frac{1}{\alpha D}\frac{(l_p^{(0)})^2}{G_0(s)}.
\label{eqn:tau_c}
\end{equation}
where
$x=x_b=\sqrt{\frac{\eta+\sqrt{\eta^2+14}}{7}sl}$ and $\xi$ is the adjustable parameter we introduced to account for the 
uncertainty in computing the curvature at the barrier top, i.e., 
$\sqrt{\nabla^2_xF(x)}|_{x=x_t\equiv\alpha}=\frac{\xi}{\alpha}\sqrt{k_BT}$. 
Note that the structure of Eq.(\ref{eqn:tau_c}) is identical to 
our previous estimate of tertiary contact formation 
time used to interpret kinetics of loop formation in proteins 
$\tau(n)\approx\frac{\langle R^2_n\rangle}{D_0P(n)}$ where $n$ is the loop length, $\langle R_n^2\rangle$ is the 
mean square distance between the two residues, $D_0$ is an effective monomer diffusion constant, 
and $P(n)$ is the loop formation probability.\cite{CamachoPNAS95,GuoBP95,ThirumalaiJPCB99}
The differences between the two lie in the numerical prefactor inside 
$\{\ldots\}$. 
In addition, in the dimensional argument used to obtain $\tau(n)$ we used $\langle R_n^2\rangle$ instead of $l_p^2$ that arises in the present theory.

Fig.\ref{fpt_s.fig} shows that the estimates of looping time using Eq.(\ref{eqn:tau_c}) and the results of 
simulations for 
the same set of parameters are in excellent agreement when $\xi\approx 7.3$ 
(see the next section for details of the simulations).
First, $\tau_c$ increases and converges to the finite value with the increasing size of end tails (decreasing $s/l$) and 
this trend manifests itself as the chain gets \emph{stiffer} and \emph{shorter} (small $s$) (see Refs.\cite{PericoMacro90} and \cite{WinnikMacro97}). 
The inset shows that, at $s=3$, $\tau_c$ increases by a factor of $\sim1.5$ when the 
total contour length of the dangling end is 5 times longer 
than the contour length of the loop.
Second, $\tau_c$ is a minimum ($\tau_c^{min}$) 
when the contour length between 
loop formation sites, $|s_1-s_2|$, is around $(3\sim 4)$ $l_p^{(0)}$ and 
$\tau_c^{min}$ shifts towards the large $s$ value 
with the increasing size of dangling ends. 
Note that when the loop size becomes large ($s>6$) 
$\tau_c$ does not depend on the length of the dangling ends. 
In non-interacting Gaussian chains the equilibrium distribution of 
any two segment along the chain is always Gaussian. 
In this case, the presence of 
the dangling ends does not affect the chain statistics.\\

\section{Simulation of loop closure dynamics}

To check the validity of the theoretical estimates for loop closure time 
we performed simulations using a coarse-grained model for ds-DNA.
The simulation procedure is identical to the one used by 
Podtelezhnikov and Vologodskii (see details in Ref.\cite{VologodskiiMacro00}).
Because the time scale of $\sim ms$ is computationally difficult to accomplish, even using Brownian dynamics (BD), 
we use a coarse-grained model of ds-DNA by choosing the pitch of the helix (10 base pairs with the diameter $l_0=3.18nm$) as a building block of a ds-DNA chain.
The energy for a worm-like chain, that is appropriate for ds-DNA, is taken to be the sum of the bending rigidity ($E_b$) term and the chain connectivity ($E_s$) term,
which respectively are given by
\begin{equation}
E_b=\alpha RT\sum_{i=1}^{N-1}\theta^2_i
\label{eqn:sim1}
\end{equation}
and
\begin{equation}
E_s=\frac{\beta RT}{l_0^2}\sum_{i=1}^N(l_i-l_0)^2.
\label{eqn:sim2}
\end{equation}  
where $T$ is the temperature, $R$ is the gas constant, $\theta_i$ is the i-th bond angle, $l_i$ is the i-th bond length. 
By choosing the parameters $\alpha=7.775$ and $\beta=50$ one can get the typical persistence length of $50nm$ for ds-DNA.

Despite the simplification in the energy function computation of the looping time through direct BD simulation still is prohibitively difficult. 
From Eq.(\ref{eqn:tau_c}) it is clear that the loop formation time can be even up to $\mathcal{O}(1)$ $sec$ for certain values of 
$l_p$ and $L$. 
To overcome this problem, Podtelezhnikov and Vologodskii used the relation between the equilibrium probability of loop formation 
and the loop closure and opening times, 
\begin{equation}
P(r_0;|s_2-s_1|,L)=\frac{\tau_o}{\tau_{IC}+\tau_o}.
\label{eqn:volo}
\end{equation}
$\tau_o$ is the loop opening time of the closed loop. 
In general $\tau_o\ll \tau_{IC}$.
This observation enables us to perform direct BD simulation for the loop dissociation rather than loop closure.
Since $P(r_0;|s_2-s_1|,L)$ is normally very small for small $r_0$ ($r_0=5nm$), there is a sampling problem. 
However, $P(r_0;|s_2-s_1|,L)$ can be found using the Markov relation 
\begin{equation}
P(r_0;|s_2-s_1|,L)=\prod_{i=1}^{n-1}P(r_i|r_{i+1})
\end{equation}
where $P(r_i|r_{i+1})$ is the conditional probability that conformations with $r<r_i$ in the subset of 
conformations with $r<r_{i+1}$. 
To obtain $P(r_i|r_{i+1})$ we performed ($n-1$)-Monte Carlo samplings using the 
pivot algorithm \cite{BishopJCP91} by iteratively adjusting the interval of the end-to-end (or interior-to-interior) 
distance $r_0<r_1<\cdots<r_n$ such that 
$P(r_i|r_{i+1})\sim 0.2$.

The results of our simulations for $P(r_0;L)$ and $\tau_o$ for both end-to-end contact and contact between segments are shown in Fig.\ref{P0_diss.fig}.
Note that there is minor difference between the contact probabilities of the end-to-end and 
of the interior-to-interior ($|s_2-s_1|/L=0.5$) segments whereas the loop opening dynamics for the chain with dangling ends 
is slower than the case without dangling ends by about $\sim$50\%.
The independence of $P(r_0;|s_2-s_1|,L)$ justifies the approximation, $G_0(s)\rightarrow G(0,s,l)$, used in 
obtaining Eq.(\ref{eqn:tau_c}).
The values of $\tau_c$ and $\tau_{IC}$  can be computed knowing  $P(r_0;|s_2-s_1|,L)$ and $\tau_o$. The results, which are shown in Fig.\ref{fpt_s.fig}, are in excellent agreement with theory.\\

\section{Interior looping dynamics in semiflexible chain with variable persistence length}

In many cases stiffness of the loop, which is involved in interior looping, is different from the overall persistence length of the chain. 
A simple example is the formation of a $\beta$-hairpin in peptides. 
In this case, the stiffness of the loop is typically less than the $\beta$-strands. 
If the $\beta$-hairpin-forming polypeptide chain is treated as a WLC then it is characterized by three fragments, namely, the loop with a persistence length 
$l_{p2}$ and the strands whose persistence lengths are $l_{p1}$ and $l_{p3}$. 
Such a variable persistence length WLC is also realized in the DNA-RNA-DNA construct in which $l_{p2}\approx 10$ \AA\ and $l_{p1}\approx l_{p3}\approx 500$ \AA. 
Constructs consisting of three WLC fragments are also used routinely in laser optical tweezer experiments. 

The Kramers type theory, used to calculate interior looping in WLC with uniform $l_p$, can be adopted to compute $\tau_{IC}$ in WLC with variable $l_p$. 
The mean field equilibrium distribution function $P(r)$ (with $r=R_{12}/|s_2-s_1|$), 
which is needed to calculate $\tau_{IC}$ is (see Appendix III for the derivation)
\begin{equation}
P(r)=\frac{4\pi Cr^2}{1-\frac{L_2}{L}r^2}\exp{\left(-\frac{3t}{4}\frac{1}{1-\frac{L_2}{L}r^2}\right)}
\end{equation}
where $L_2=|s_2-s_1|$ is the contour length of the loop part of WLC with persistence length $l_{p2}$, $L$ is the total contour length of the 
chain, $t=L/l_0^{eff}$. The effective persistence length of the WLC, consisting of three segments $0<s<s_1$ with $l_{p1}$,  $s_1<s<s_2$ with $l_{p2}$, 
and $s_2<s<L$ with $l_{p3}$ is 
\begin{equation}
l_p^{eff}=\left(\sum_{i=1}^3\frac{L_i}{L}\frac{1}{\sqrt{l_{pi}}}\right)^{-2}.
\label{eqn:eff_lp}
\end{equation}
In the mean field approximation, the WLC in which $l_p$ varies along the contour in a discrete manner, is equivalent to a WLC with an 
effective persistence length. 
It follow from Eq.\ref{eqn:eff_lp} that the effective persistence length is determined by the smallest $l_{p_i}$.

Consider the simplest case $L_2=L_{int}$, $L_1=L_3=\frac{L-L_{int}}{2}$, $l_{p1}=l_{p3}=l_p^H$, and $l_{p2}=l_p^{(0)}$. 
In this case, 
\begin{equation}
\frac{l_p^{eff}}{l_p^{(0)}}=\frac{1}{\left[(1-x)\sqrt{\frac{l_p^H}{l_p^{(0)}}}+x\right]}
\label{eqn:lp_eff}
\end{equation}
where $x=L_{int}/L$. 
It follows from Eq.\ref{eqn:lp_eff} that if $l_p^H>l_p^{(0)}$ (handle is stiffer than the loop) then $l_p^{eff}>l_p^{(0)}$. 
Because in the Kramers description $\tau_{IC}$ is controlled by $l_p^{eff}$ we expect that interior looping time is 
greater than $\tau_{IC}$ for a chain with uniform $l_p$. 
In the opposite limit, $l_p^H<l_p^{(0)}$, (handle is softer than the loop) $l_p^{eff}<l_p^{(0)}$. 
Consequently, attaching a soft handles should enhance the rate of interior looping. 

The interior looping kinetics for a WLC copolymer for different values of the loop and handle persistence lengths are 
shown in Fig.\ref{fpt_hetero.fig}. 
In accord with the arguments given above, we find that when $l_p^H/l_p^{(0)}=2$ (stiff handles) $\tau_{IC}$ increases substantially 
compared to $\tau_{IC}(\equiv\tau_{IC}^o)$ for $l_p^H/l_p^{(0)}=1$ for all values of $|s_2-s_1|/l_p^{(0)}$ (Fig.\ref{fpt_hetero.fig}). 
Similarly, when the handle is softer than that of the loop, $\tau_{IC}$ decreases appreciably compared to $\tau_{IC}^o$ which is the interior looping 
the case when the chain has uniform stiffness. 
In the interesting regime of $|s_2-s_1|/l_p^{(0)}\approx (2-4)$ we predict a dramatic increase in $\tau_{IC}$ compared to $\tau_{IC}^o$ 
when $l_p^H/l_p^{(0)}>1$ and a substantial decrease in $\tau_{IC}/\tau_{IC}^o$ when $l_p^H/l_p^{(0)}<1$ (see inset in Fig.\ref{fpt_hetero.fig}). 
Thus, stiff handles retard interior loop kinetics whereas soft handle enhances rates of interior loop formation.   

\section{Effect of monomer-monomer interaction on loop closure times}
 
Majority of the recent experiments on dynamics of loop formation have been analyzed using simple polymer 
models that do not explicitly consider interaction between monomer segments. 
In a number of cases there are physical interactions between monomers. 
For instance, DNA is charged and the interaction between the monomers can be approximately described using the short ranged 
Debye-H{\"u}ckel potential. 
Similarly, solvent-mediated interactions also arise especially when considering proteins. 
For these reasons it becomes necessary to consider an 
interplay between chain stiffness, entropic fluctuations of the polymer and nonlocal interaction between monomer segments. 

The non-linear problem, that arises from the constraint $\textbf{u}^2(s)=1$, in a non-interacting semiflexible 
chain is further exacerbated when interactions between monomers are taken into account. \cite{LeeEJB99} 
To circumvent this problem we assume that the effect of intra-chain interaction is to only alter the 
effective persistence length. 
We compute the loop closure kinetics using Eq.(\ref{eqn:tau_c}) with a renormalized persistence length that explicitly depends on the 
nature of the interaction between the monomers. 
This approximation is in the same spirit as the local equilibrium assumption used in this study. 

To calculate the renormalized persistence length $l_p^{(R)}$ 
we follow the procedure due to Hansen and Podgornik \cite{PodgornikJCP01} 
who used a mean-field weight (similar to $\Psi_{MF}(\textbf{u}(s))$ in the presence of non-local interaction, 
$V(\textbf{r}(s)-\textbf{r}(s'))$ between monomers $s$ and $s'$. 
The standard field-theoretic procedure is to use the Hubbard-Stratonovich transformation via auxiliary fields to 
eliminate the non-Markovian nature of $V(\textbf{r}(s)-\textbf{r}(s'))$. 
Using stationary phase approximation to evaluate the optimal values of the auxiliary fields 
they \cite{PodgornikJCP01} obtained an 
expression for $l_p^{(R)}$ for arbitrary potential $V(\textbf{r}(s)-\textbf{r}(s'))$. 
In our applications, we assume that the charged monomers interact via the screened Coulomb interaction 
$V(r)=\frac{k_BTl_B}{A^2}\frac{e^{-\kappa r}}{r}$ where 
$l_B$ is the Bjerrum length ($e^2/4\pi\epsilon_0k_BT=l_B$), 
and $A$ is the effective separation between charges on the monomer.
 The use of Debye-H{\"u}ckel potential is appropriate when considering ds-DNA in monovalent (Na$^+$) counterions. 
With this choice of $V(r)$ 
the renormalized persistence length becomes $l_p^{(R)}=l_p^{(0)}+\delta l_p^{(R)}$, and is given by \cite{PodgornikJCP01}
\begin{equation}
l_p^{(R)}=l_p^{(0)}+\overline{V}_1(l_p^{(R)})^2 I(L/l_p^{(R)},\xi)
\label{eqn:renormalization}
\end{equation}
with
\begin{eqnarray}
I(L/l_p^{(R)},\overline{\xi})=\int^{L/l_p^{(R)}}_{0}dzz^4e^{-\sqrt{\overline{B}}/\overline{\xi}}\left(\frac{1}{\overline{\xi}\overline{B}}+\frac{1}{\overline{B}^{3/2}}\right)
\label{eqn:Integral}
\end{eqnarray}
where $\overline{V}_1=l_B/(12\sqrt{2}dA^2)$, $\overline{B}(z)=z-1+e^{-z}$, 
and $\overline{\xi}=1/(\sqrt{2}\kappa l_p^{(R)})$.
The integral $I(L/l_p^{(R)},\overline{\xi})$ 
has different asymptotic behavior depending on the two parameters $\kappa l_p^{(R)}$ and $L/l_p^{(R)}$. 
(i) If $l_p^{(R)}>\kappa^{-1}$ ($\overline{\xi}<1$), i.e., the persistence length is greater than the screening length, then the contribution due to electrostatic 
interaction can be treated perturbatively. In this case the upper limit of the integral in Eq.(\ref{eqn:Integral}) is effectively set to infinity. 
We find $\overline{B}(z)\approx z^2$ and  
$I(L/l_p^{(R)},\overline{\xi})\sim \overline{\xi}^2$, which is also small since $\overline{\xi}<1$.
Therefore, $\delta l_p^{(R)}\sim \frac{l_B}{\kappa^2 A^2}$, which coincides with the OSF result.\cite{OdijkJPS77,SkolnickMacro77}
For electrostatic contribution to persistence length of a polyelectrolyte chain the limit $l_p^{(R)}> \kappa^{-1}$ is most appropriate for DNA .
(ii) if $l_p^{(R)}<\kappa^{-1}$ ($\overline{\xi}>1$), i.e., the persistence length is 
smaller than the screening length, there is substantial interaction between 
the chain segments beyond the length scale of 
$l_p^{(R)}$.
We believe this situation is difficult to be realized in experiments involving biopolymers. 
In this case, the integral up to $z=\overline{\xi}^2$ becomes important,
$\overline{B}(z)\sim z$ and $I(L/l_p^{(R)},\overline{\xi})\sim\overline{\xi}^7$. 
Therefore, $\delta l_p^{(R)}\sim \frac{l_B}{\kappa^7 A^2}$.

To calculate loop closure times 
the renormalized persistence length $l_p^{(R)}$ is numerically computed for 
each parameter set 
(contour length $L$, inverse screening length $\kappa^{-1}$) 
and we use $\tau_c$ (Eq.(\ref{eqn:tau_c})) with $l_p^{(0)}\rightarrow l_p^{(R)}$. 
For the ds-DNA in the monovalent salt solution (concentration $c$) the parameters in the semiflexible 
chain model are $l_B=7.1$\AA, $A=1.7$\AA\ and $\kappa=\sqrt{8\pi l_Bc}$. 
The results for $\tau_c$ and $\tau_{IC}$ are plotted in Figs.\ref{fpt_conc.fig}, \ref{fpt_s_combo.fig}.
First, the cyclization times are computed 
as a function of $L$ at various salt concentrations (Fig.\ref{fpt_conc.fig}). 
We find that $\tau_c$ shows a dramatic increase as $c$ is varied at small values of 
$L/l_p^{(0)}$. 
The electrostatic repulsion retards loop closure times, as the salt concentrations 
(strong inter-segment repulsion) decrease (Fig.\ref{fpt_conc.fig}).
Because of the interplay between bending rigidity and chain entropy $\tau_c$ has a minimum at $t=t^*$. The value 
of $t^*$ shifts from $t^*=3$ to $t^*=6$ as $c$ decreases. 
The inset in Fig.\ref{fpt_conc.fig} shows that 
there is practically no change in $\tau_c$ at $t=3$  
if $c\gtrsim 100$ $mM$, which is near the physiological concentration ($150$ $mM$ $\mathrm{Na}^+$). 
In this range of $c$ the electrostatic contribution to the persistence length is small so that 
$l_p^{(R)}$ is 
almost the same as $l_p^{(0)}$. 
Note that the $c$ corresponding to the condition $l_p\kappa=1$ for $l_p=50$ $nm$ is $c\approx 40$ $mM$.

The dependence of $\tau_{IC}$, which examines the effect of the dangling ends, at high and low concentrations and various values of $|s_2-s_1|/L$ on $|s_2-s_1|$ is shown in Fig.\ref{fpt_s_combo.fig}. 
The insets of Fig.\ref{fpt_s_combo.fig} show an increased time scale 
at low salt concentration (10 $mM$) compared with high salt concentration (500 $mM$). 
When $\tau_{IC}$ for $|s_2-s_1|/L=0.2$ is compared with  $|s_2-s_1|/L=1$, 
$\tau_c$ increases by a factor of $\sim2.7$ at $c=10$ $mM$ 
whereas the increase is about a factor of $\sim 1.5$ for $c=500$ $mM$.
The effect of dangling ends on loop formation dynamics manifests itself more clearly 
at low salt concentration when electrostatic repulsion is prominent and at small ratio of $|s_2-s_1|/l_p^{(0)}$.

\section{Dependence of $l_p^{(R)}$ on the monomer-monomer interaction potential} 
In our theory $\tau_c$ and $\tau_{IC}$ can be determined provided $l_p^{(R)}$ and the distance distribution functions are known. 
To examine the variation of $l_p^{(R)}$ on the nature of monomer-monomer interactions we have computed $l_p^{(R)}$ for different potentials $V(r)$. 
Sets of equilibrium conformations of 50-mer bead-spring model (see Eq.(\ref{eqn:sim1}),(\ref{eqn:sim2})) 
are generated with 
different bending rigidity, $\kappa_b$, and with different non-local potentials 
$V(r)=1/r^{\alpha}$($\alpha=$1, 2, 4, 6, 12). 
In each case the effective persistence length $l_p^{(R)}$ is 
computed by $\frac{1}{1-\cos{\langle\theta\rangle}}$ \cite{LeeBJ04} where $\langle\theta\rangle$ is 
the ensemble average of the angle formed by three consecutive beads. 
We show the simulated radial distribution function and the effective persistence lengths for 
different $\kappa_b$ values for the various ranges of nonlocal interaction 
in Fig.\ref{interaction_pr.fig}-(a),(b).  
The results from radial distribution function (Fig.\ref{interaction_pr.fig}-(a)) 
and the persistence length show 
that $l_p^{(R)}\approx l_p^{(0)}$ when the interaction is short-ranged, i.e., $\alpha>4$. 
When $\alpha<4$ then the effective interaction between monomers leads to an increase in the persistence length (Fig.\ref{interaction_pr.fig}-(b)). These results are consistent with 
the field theoretical approach by Hansen and Podgornik.\cite{PodgornikJCP01}
Considering that the excluded volume potential is of short range nature (modeled using $\sim r^{-12}$ or 
delta function) we conclude that the excluded volume effect on the looping dynamics of rigid polymer chain is 
negligible. 
Note that the screened electrostatic potential $V(r)=\frac{e^{-\kappa r}}{r}$ 
can be either a short or a long range potential depending on the value of $\kappa$.   

These calculations, especially changes in $l_p$ as the range of interaction is altered, explain the reason that a simple WLC model works 
remarkably well in a number of applications. 
For example, the response of DNA, RNA,\cite{Bustamante2,Bustamante4} and proteins \cite{GaubSCI97} to mechanical force has been routinely analyzed using WLC. 
Surprisingly, recent analysis of 
small angle x-ray scattering measurements \cite{CaliskanPRL05} on ribozymes 
have shown that the distance 
distribution function can be quantitatively fit using $P(R_e)$. 
In these biopolymers the interactions that determine the conformations are vastly different. 
However, the results in Fig.\ref{interaction_pr.fig}-(b) show that as long as these effective interactions are short-ranged $l_p$ should not differ from the bare persistence length. 
This key result rationalizes the use of WLC in seemingly diverse set of problems. 
\\

\section{Conclusions}

In this paper we have used theory and explicit simulations of worm-like chains 
to examine loop formation dynamics with emphasis on kinetics
of contact formation between monomers that are in the interior of the chain.
The Kramers theory, adopted to describe looping time scales using the
analytically computed potential of force between the contacting (or reacting)
groups, gives results that are in quantitative agreement with simulations. The
theory \cite{HaELett03} for $\tau_{IC}$ and $\tau_c$ contains one parameter that was 
introduced to account for the uncertainty in the estimate of the frequency
at the transition state (Fig.\ref{Dfunc.fig}) in the intramolecular reaction (Fig.\ref{looping.fig}). 
The present study also provides a justification for the use of Kramers-like
theory in describing looping dynamics by explicit comparison with simulations
of semiflexible chains.  Although several questions of fundamental theoretical
importance remain previous studies, beginning with the pioneering work by WF,
and the present study have given a practical analytic formula to analyze
most of the recent experimental data on proteins and DNA.  We conclude the 
paper with a few additional comments.

\begin{enumerate}

\item The present work and several previous studies,\cite{HofrichterJPCB02,VologodskiiMacro00,HaELett03,CamachoPNAS95} which have examined the
effect of stiffness on looping dynamics, have shown that the rates of
cyclization and interior looping must slow down as the loop length becomes
small. In other words, there must be a turnover in the plot of $k_{\alpha}$
($\alpha = c$ or $IC$) as $s$ decreases (see Fig.\ref{fpt_s.fig} in which $\tau_{IC}$ as 
a function of $s$ is shown).  For the parameters used in Fig.\ref{fpt_s.fig} the 
turnover occurs around $s \approx (3-4)$.  Such a crossover has been observed
in the cyclization kinetics of DNA and in simulations of worm like chain
models.  When $s$ is small the time scales for loop formation can be
substantially large ($\sim\mathcal{O}(1)$ $sec$). 

The effect of stiffness on cyclization rate in disordered peptides has also 
been emphasized.\cite{HofrichterJPCB02}
For the construct $Cys-(Ala-Gly-Gln)_{j}-Trp$ with
$j$ from $1-6$ the stiffness effects are evident at $j \approx 3$.\cite{HofrichterJPCB02}
However, these
authors did not observe the theoretically predicted turnover in this construct for which the persistence length is estimated to be 
$l_p \approx 0.7nm$. Using the results in Fig. (3) we
predict that the turnover must occur only when the number of peptide bonds is
less than about 3. 
This limit has not been reached in the experiments by 
Lapidus \emph{et. al.} \cite{HofrichterJPCB02}
For the construct $(Gly-Ser)_j$ Hudgins \emph{et. al.} \cite{HudginsJACS02} have
clearly observed a turnover when the number of peptide decreases below about
4.  The observation of Hudgins \emph{et. al.} is consistent with our prediction that 
turnover in cyclization rates in disordered polypeptides occurs when $s\lesssim 3$.
When the number of residues in the polypeptides chain becomes too small then measuring $\tau_c$ 
using bulky donor-acceptor pairs in FRET experiments is difficult. 
In this situation other methods \cite{GrayPNAS03} could be used. 

\item For the parameters used in Fig. (3) the difference between $\tau_c$ and
$\tau_{IC}$ is no more than about a factor of four. However, if charged 
interactions between monomers become relevant then $\tau_{IC}$ can be
very different from $\tau_c$. At both low and high values of the
salt concentration the $\tau_c$ and $\tau_{IC}$ can differ by nearly
an order of magnitude (see Fig.(6)).  These variations are significant because
$\tau_c$ in polypeptide chains studied thus far varies by less than a factor of ten as the
number of residues is varied from 5 to 20.  It would be interesting
to probe looping dynamics by varying the net charge on polypeptides. We should
also stress that as the salt concentration increases the electrostatic
interactions in a high dielectric medium are effectively short ranged. In
this case $\tau_{IC}$ is determined essentially by the bending rigidity of
the backbone (Fig.\ref{interaction_pr.fig}).

\item The Kramers based theory for $\tau_c$ and $\tau_{IC}$ is a convenient way to measure persistence length of polypeptide chains as 
a function of temperature and denaturant concentration. 
Recent measurements suggest that $l_p$ depends on urea concentration.\cite{KiefhaberJMB05}
More importantly, there appears to be strong sequence effects in $\tau_c$ which, at the level of polymer-based theories, must reflect 
changes in $l_p$. For example, $\tau_c$ for polyproline deviates substantially from ideal chain behavior.\cite{NauRCI05}
Similar measurements of $\tau_c$ for other polypeptides along with the simple theory can be used to extract how $l_p$ varies with sequence. 

\item The dependence of interior looping time on the ratio of the persistence lengths of loop and the handle shows that in the 
interesting range of $s\sim(2-4)$ $\tau_{IC}$ can be substantially larger than $\tau_{IC}^o$ for a chain in which 
$\l_p^H/l_p^{(0)}=1$ (see inset in Fig.\ref{fpt_hetero.fig}). 
This case is directly applicable to $\beta$-hairpin formation that is controlled by formation of a loop with persistence length that is 
less than that of the strands.\cite{KlimovPNAS00}
From Fig.\ref{fpt_hetero.fig} it follows that as the stiffness of the loop increases the interior looping time also increases. 
This conclusion is in accord with explicit simulations of coarse-grained models of $\beta$-hairpin formation that showed that enhancement of loop 
stiffness retards rate of $\beta$-hairpin formation.\cite{KlimovPNAS00}

\end{enumerate}


\numberwithin{equation}{section}
\section*{APPENDIX A}
In this Appendix we outline the steps leading to Eq.(\ref{eqn:functional_result}). 
The distribution function $G(R_{12},s_1-s_2)$ is 
\begin{align*}
G(R_{12},s_1-s_2)&=e^{\lambda L+2\delta}\int\frac{d^3\textbf{k}}{(2\pi)^3}\int d\textbf{u}_0d\textbf{u}_{s_1}d\textbf{u}_{s_2}d\textbf{u}_L\nonumber\\
&\times e^{-\delta\textbf{u}_0^2}Z(\textbf{u}_0,\textbf{u}_{s_1};s_1)\nonumber\\
&\times e^{-\frac{\textbf{k}^2}{4\lambda}|s_2-s_1|+i\textbf{k}\cdot \textbf{R}_{12}}Z(\textbf{u}_{s_1}+\frac{i\textbf{k}}{2\lambda},\textbf{u}_{s_2}+\frac{i\textbf{k}}{2\lambda};s_2-s_1)\nonumber\\
&\times e^{-\delta\textbf{u}_L^2}Z(\textbf{u}_{s_2},\textbf{u}_L;L-s_2)
\label{eqn:A1}
\tag{A.1}
\end{align*}
Using the expression Eq.(\ref{eqn:Z}) for $Z(\textbf{u}_s,\textbf{u}_0,s)$ and carrying out the integrals over the $\textbf{u}$ variables Eq.(\ref{eqn:A1}) becomes 
\begin{align*}
G(R_{12} &,s_1-s_2)=e^{\lambda L+2\delta}\left(\frac{\pi\sinh{\Omega s_1}}{\Omega_p}\right)^{-3/2}\left(\frac{\pi\sinh{\Omega(s_2-s_1)}}{\Omega_p}\right)^{-3/2}\left(\frac{\pi\sinh{\Omega(L-s_2)}}{\Omega_p}\right)^{-3/2}\nonumber\\
&\times \int\frac{d^3\textbf{k}}{(2\pi)^3}\exp{\left(i\textbf{k}\cdot\textbf{R}_{12}-\frac{\textbf{k}^2}{4\lambda}|s_2-s_1|+\frac{\textbf{k}^2}{2\lambda^2}\Omega_p(\coth{\Omega(s_2-s_1)}-\frac{1}{\sinh{\Omega(s_2-s_1)}})\right)}\nonumber\\
&\times \left(\frac{\pi}{\det\mathcal{A}}\right)^{3/2}\exp{\left(\frac{1}{4}\textbf{b}^T(\textbf{k})\mathcal{A}^{-1}\textbf{b}(\textbf{k})\right)}
\tag{A.2}
\end{align*}
where
\begin{align*}
\begin{footnotesize}
\mathcal{A}=\left(\begin{array}{cccc}
\delta+\Omega_p\coth{\Omega s_1} & -\frac{\Omega_p}{\sinh{\Omega s_1}} & 0 & 0\\
-\frac{\Omega_p}{\sinh{\Omega s_1}} &\Omega_p(\coth{\Omega s_1}+\coth{\Omega(s_2-s_1)})& -\frac{\Omega_p}{\sinh{\Omega(s_2-s_1)}} & 0\\
0&-\frac{\Omega_p}{\sinh{\Omega(s_2-s_1)}} &\Omega_p(\coth{\Omega(s_2-s_1)}+\coth{\Omega(L-s_2)})& -\frac{\Omega_p}{\sinh{\Omega(L-s_2)}}\\
0& 0&  -\frac{\Omega_p}{\sinh{\Omega(L-s_2)}}&\delta+\Omega_p\coth{\Omega(L-s_2)}\end{array}\right), 
\end{footnotesize}
\tag{A.3}
\end{align*}
and 
\begin{align*}
\textbf{b}^T=\frac{i\textbf{k}}{\lambda}\Omega_p(\coth{\Omega(s_2-s_1)-\frac{1}{\sinh{\Omega(s_2-s_1)}}})\left(\begin{array}{cccc}
0&1&1&0\\
\end{array}\right).
\tag{A.4}
\end{align*}
The integration with respect to $\textbf{k}$ leads to 
\begin{align*}
G(R_{12},s_1,s_2)&=e^{\lambda L+2\delta}\left(\frac{\pi\sinh{\Omega s_1}}{\Omega_p}\right)^{-3/2}\left(\frac{\pi\sinh{\Omega(s_2-s_1)}}{\Omega_p}\right)^{-3/2}\left(\frac{\pi\sinh{\Omega(L-s_2)}}{\Omega_p}\right)^{-3/2}\nonumber\\
&\times\left(\frac{\pi}{\det\mathcal{A}}\right)^{3/2}\frac{1}{(2\pi)^3}\left(\frac{4\pi\lambda^2}{Q(s_1,s_2;\lambda,\delta)}\right)^{3/2}e^{-\frac{\lambda^2R_{12}^2}{Q(s_1,s_2;\lambda,\delta)}}
\tag{A.5}
\end{align*}
where 
\begin{align*}
\det\mathcal{A}=\frac{\Omega_p^2\sinh{\Omega L}}{\sinh{\Omega s_1}\sinh{\Omega (s_2-s_1)}\sinh{\Omega(L-s_2)}}\{(\delta^2+\Omega_p^2)+2\delta\Omega_p\coth{\Omega L}\},
\tag{A.6}
\end{align*}
\begin{align*}
Q(s_1,s_2;\lambda,\delta)=|s_2-s_1|\lambda-2\Omega_p(\coth{\Omega(s_2-s_1)}-\frac{1}{\sinh{\Omega(s_2-s_1)}})+M(s_1,s_2),
\tag{A.7}
\end{align*}
and
\begin{align*}
M(s_1,s_2;\lambda,\delta)&=\frac{\Omega_p\sinh{\frac{\Omega(s_2-s_1)}{2}}\tanh^2{\frac{\Omega(s_2-s_1)}{2}}}{2\delta\Omega_p\cosh{\Omega L}+(\delta^2+\Omega^2_p)\sinh{\Omega L}}\nonumber\\
&\times\{4\delta\Omega_p\cosh{\frac{\Omega(2L-(s_2-s_1))}{2}}+2(\delta^2+\Omega_p^2)\sinh{\frac{\Omega(2L-(s_2-s_1))}{2}}\nonumber\\
&-(\delta^2-\Omega^2_p)(\sinh{\frac{\Omega(2L-3s_1-s_2)}{2}}-\sinh{\frac{\Omega(2L-s_1-3s_2)}{2}})\}
\tag{A.8}
\end{align*}
In limit $L\gg s_2\gg s_1\gg 1$, $M(s_1,s_2)\rightarrow\Omega_p$ and $Q(s_1,s_2;\lambda,\delta)\rightarrow|s_2-s_1|\lambda-\Omega_p$. 
As a result of translational symmetry along the chain $G(R,s_1,s_2)=G(R,|s_2-s_1|)$. 
Using $\mathcal{F}(\lambda,\delta)\approx -\ln{G(R_{12},s_1,s_2)}$ leads to Eq.(\ref{eqn:functional_result}). 

\numberwithin{equation}{section}
\section*{APPENDIX B}
In obtaining the stationarity condition to evaluate $\lambda$ and $\delta$ we first took the thermodynamic limit ($L\rightarrow\infty$) and then 
calculated the optimal values of $\lambda$ and $\delta$. 
It is technically necessary to solve the stationarity condition $\frac{\mathcal{F}}{\partial\lambda}=\frac{\partial\mathcal{F}}{\partial\delta}=0$ 
before taking the $L\rightarrow\infty$ limit. In this appendix we examine the consequence of taking the thermodynamic limit after solving for optimal values of 
$\lambda$ and $\delta$. 
For simplicity, set $|s_2-s_1|=L$.
The variational equations for $\lambda$ and $\delta$ become
\begin{equation*}
\delta+\Omega_p=\frac{3}{2}
\tag{B.1}
\end{equation*}
and 
\begin{equation*}
L\left[\left(\frac{3}{4\Omega_p}-1+\frac{R^2}{L^2}\right)-\frac{l_p}{L}\left(\frac{9}{8\Omega_p^2}-\frac{1}{2\Omega_p}\right)\right]=0
\tag{B.2}
\end{equation*}
From the second relation we find two roots for $\Omega_p$, namely, 
\begin{equation*}
\Omega_p^{\pm}=\sqrt{\frac{\lambda^{\pm}l_p}{2}}=\frac{3(1+\frac{1}{t})}{4(1-r^2)}\left[\frac{1}{2}\pm\frac{1}{2}\sqrt{1-\frac{12(1-r^2)}{t(1+\frac{1}{t})^2}}\right]
\tag{B.3}
\label{eqn:B3}
\end{equation*}
where $t=L/l_0$, $l_0=\frac{2}{3}l_p$ and $r=R/L$.
There are no restrictions on the values of $L$ and $l_p$ in Eq.(\ref{eqn:Functional}) in which the thermodynamic limit is taken first. 
However, when the order of operation is exchanged there is a possibility 
that the two roots are $\Omega_p^{\pm}$ that can be imaginary. 
For $\Omega_p^+$ we retrieve the same stationary phase condition as Eq.(\ref{eqn:stationarycondition}) only if $L\gg l_p$($t\gg 1$)
The second root $\Omega_p^-=0$ but this can be discarded since $\lambda\neq 0$. 
Although there are multiple saddle points, we can always deform the contour such that the contour passes the saddle point with $\mathcal{F}'(\Omega_p^+)=0$ 
which satisfies the stationarity condition. 
In addition, $\frac{1}{12}(10-t-\frac{1}{t})<r^2<1$ should be always satisfied for $\Omega_p$ to be real (Eq.(\ref{eqn:B3})).\\

\numberwithin{equation}{section}
\section*{APPENDIX C}

We calculate $P(\textbf{R}_{12},|s_2-s_1|)$ for a semiflexible chain in which the persistence length of the loop is different from that of the dangling ends. 
The chain can be thought of as a triblock WLC copolymer which is an appropriate model for RNA hairpins or $\beta$-hairpins 
in disordered polypeptides. 
The interior $\textbf{R}_{12}$ distribution function is  
\begin{align}
  G(\textbf{R}_{12};L_1,L_2,L_3)&=\langle\delta(\textbf{R}_{12}-\int_{L_2}\textbf{u}(s)ds)\rangle_{MF}\nonumber\\
  &=\frac{\int D[\textbf{u}(s)]\delta(\textbf{R}_{12}-\int_{L_2}\textbf{u}(s)ds)\Psi_{MF}[\textbf{u}(s)]}{\int D[\textbf{u}(s)]\Psi_{MF}[\textbf{u}(s)]},
\tag{C.1}
\end{align}
where $\textbf{u}(s)$ is a tangent vector at position $s$.
Suppose the chain consists of three different parts characterized by the persistence lengths ($l_{pi}$) and the contour lengths ($L_i$) with $i=1,2,3$ ($L=\sum_{i=1}^3L_i$). 
The exact weight $\Psi[\textbf{u}(s)]\varpropto\exp{[-\sum_{i=1}^3\frac{l_{pi}}{2}\int_{L_i}ds\left(\frac{\partial\textbf{u}}{\partial s}\right)^2}]\prod\delta(\textbf{u}^2(s)-1)$ is replaced by the mean field weight $\Psi_{MF}[\textbf{u}(s)]$. 
\begin{equation}
  \Psi_{MF}[\textbf{u}(s)]\varpropto
  \exp{[-\sum_{i=1}^3\frac{l_{pi}}{2}\int_{L_i}\left(\frac{\partial\textbf{u}(s)}{\partial s}\right)^2ds-\sum_{I=1}^3\lambda\int_{L_i}(\textbf{u}^2(s)-1)ds-\delta[(\textbf{u}^2_0-1)+(\textbf{u}^2_L-1)]]}.
  \tag{C.2}
  \label{eqn:weight_mix}
\end{equation}
As in Appendix A, 
the Lagrange multipliers $\lambda$ and $\delta$ are used to enforce the 
local constraint $\textbf{u}^2(s)=1$. 
Following, exactly the procedure outlined in Appendix A we find that 
in the $L\rightarrow\infty$ limit, the analogue of Eq.\ref{eqn:Functional} becomes, 
\begin{align}
\mathcal{F}[\lambda,\delta]&\approx -(L\lambda+2\delta)\nonumber\\
&+\sum_{i=1}^3\frac{3}{2}\ln\frac{e^{\Omega_i L_i}}{\Omega_{pi}}+\frac{3}{2}\ln{[(\delta+\Omega_{p1})(\Omega_{p1}+\Omega_{p2})(\Omega_{p2}+\Omega_{p3})(\delta+\Omega_{p3})]}+\frac{3}{2}\ln{\frac{L_2}{\lambda}}+\frac{\textbf{R}_{12}^2\lambda}{L_2}\nonumber\\
&=L\left[\frac{3}{2}\left(\Omega_1\frac{L_1}{L}+\Omega_2\frac{L_2}{L}+\Omega_3\frac{L_3}{L}\right)-\lambda(1-\frac{L_2}{L}\frac{\textbf{R}_{12}^2}{L_2^2})\right]\nonumber\\
&+\frac{3}{2}\ln{\left[\frac{\Omega_{p2}}{\lambda}\left(\frac{\delta}{\Omega_{p1}}+1\right)\left(\frac{\Omega_{p1}}{\Omega_{p2}}+1\right)\left(\frac{\Omega_{p3}}{\Omega_{p2}}+1\right)\left(\frac{\delta}{\Omega_{p3}}+1\right)\right]}+\frac{3}{2}\ln{L_2}-2\delta.
\tag{C.3}
\label{eqn:Functional_mix}
\end{align}
where $\Omega_i=\sqrt{2\lambda/l_{pi}}$ and $\Omega_{pi}=\Omega_il_{pi}/2$. 
The major contribution to the integral comes from the sets of $\lambda$ and $\delta$ which pass the saddle point of a stationary phase contour on $\mathbf{Re}\{\mathcal{F}\}$ plane.
The stationary condition for 
$\lambda$ by taking derivative with respect to $\lambda$ by retaining the leading term in $L$, 
which leads to 
\begin{equation}
\lambda^{1/2}=\frac{3}{4(1-\frac{L_2}{L}r^2)}\left(\sum_{i=1}^3{\sqrt{\frac{2}{l_{pi}}}\frac{L_i}{L}}\right).
\tag{C.4}
\label{eqn:lambda_value}
\end{equation}
where $r=\frac{R_{12}}{L_2}$.
Similarly, the condition for $\delta$ results 
\begin{equation}
\frac{1}{\delta+\Omega_{p1}}+\frac{1}{\delta+\Omega_{p3}}=\frac{4}{3}.
\tag{C.5}
\label{eqn:delta_value}
\end{equation}
Substituting Eq.\ref{eqn:lambda_value} and \ref{eqn:delta_value} to 
$G(R)\approx \exp{(-\mathcal{F}[\textbf{R};\lambda,\delta])}$
gives the desired distribution function, 
\begin{equation}
G(R_{12};\{l_{pi}\},\{L_i\})=\frac{N}{(1-\frac{L_2}{L}r^2)^{9/2}}\exp{\left[-\frac{9}{8}\frac{L}{1-\frac{L_2}{L}r^2}\left(\sum_{i=1}^3\sqrt{\frac{1}{l_{pi}}}\frac{L_i}{L}\right)^2\right]}.
\tag{C.6}
\end{equation}
The effective persistence length in the mean field approximation is 
\begin{equation}
l_p^{eff}=\left(\sum_{i=1}^3\frac{L_i}{L}\frac{1}{\sqrt{l_{pi}}}\right)^{-2}.
\tag{C.7}
\label{eqn:persistence}
\end{equation}
The result derived for a triblock WLC can be generalized into $N$-block WLC 
with persistence lengths $\{l_{pi}\}$ and contour lengths $\{L_i\}$. 
The distribution function is 
\begin{equation}
P(r)=\frac{4\pi Cr^2}{(1-\frac{L_2}{L}r^2)^{9/2}}\exp{\left(-\frac{3t}{4}\frac{1}{1-\frac{L_2}{L}r^2}\right)}
\tag{C.8}
\end{equation}
where $t=L/l_o^{eff}$, $l_o^{eff}=\frac{2}{3}\left(\sum_{i=1}^N\frac{L_i}{L}\frac{1}{\sqrt{l_{pi}}}\right)^{-2}$, $r=R_{12}/|s_2-s_1|$ and $C$ is a normalization constant.

{\bf Acknowledgements:} 
We are grateful to William A. Eaton for useful discussions. 
This work was supported in part by a grant from the National Science Foundation through grant number NSF CHE 05-14056. 

\newpage

\newpage

\section*{FIGURE CAPTIOINS}

\begin{figure}[ht]
\mbox{\subfigure[]{\includegraphics[width=4.00in]{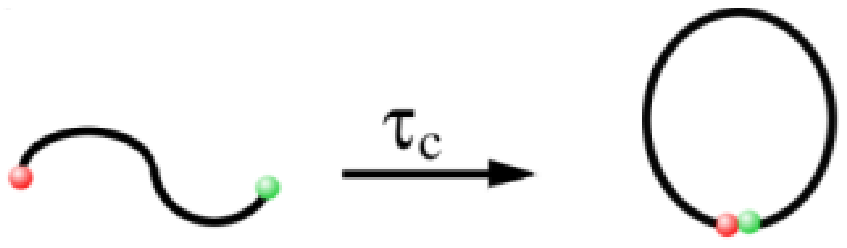}}}
\mbox{\subfigure[]{\includegraphics[width=4.00in]{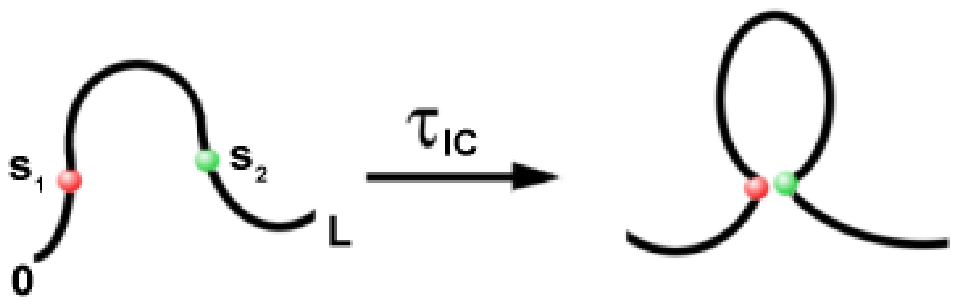}}}
\caption{\label{looping.fig}}
\end{figure}

{{\bf Figure} \ref{looping.fig} :}
Loop formation in semiflexible polymer chains: (a) Cyclization event. (b) Interior contact (IC) 
formation between monomers $s_1$ and $s_2$ in the chain interior. 
The segment length $s_1$ and $L-s_2$ are referred to either as dangling ends or handles in the text. 

{{\bf Figure} \ref{Dfunc.fig} :}
Comparison between the interior distance distribution functions for different size of the loops 
($\frac{|s_2-s_1|}{L}=$1, 0.8, 0.5). 
The value of $|s_2-s_1|/L=1$ corresponds to end-to-end distribution function. 
The flexible chain limit ($t=10$) is on the left and the right panel is for stiff chains ($t=2$). 
The panel on the right shows a sketch of the effective potential $F(r)=-k_BT\log{P(r)}$ for the case 
$|s_2-s_1|/L=0.8$. 
The wells at $R_b$ and the barrier top $R_t$ are highlighted.

{{\bf Figure} \ref{fpt_s.fig} :}
Plots of loop formation time ($\tau_{IC}$) obtained using Eq.(\ref{eqn:tau_c}) 
as a function of the distance 
between site $s_1$ and $s_2$ expressed in terms of $l_p^{(0)}$ 
for various size of dangling ends expressed by 
$s/l(=|s_1-s_2|/L)$.
Here $l=L/l_p^{(0)}$ and $s=|s_2-s_1|/l_p^{(0)}$. 
The specific values for parameters are $l_p^{(0)}=50$ $nm$, 
$D=2D_0=1.54\times 10^{-11}m^2/s$ and $\alpha=0.1$. 
The insets are for
the loop closing time at $|s_1-s_2|/l_p^{(0)}=3$ as a function of $s/l$. 
For two sites separated by $|s_1-s_2|$ along the chain the longer loop closing time is expected if the sites of interest are connected by long dangling ends. 
If the separation is much larger than the 
persistence length ($|s_1-s_2|>>l_p^{(0)}$) the effect of dangling end on 
loop closing time vanishes.
The results of simulations for the same set of parameters are shown in symbols. The excellent agreement between theory and simulations validates the assumptions leading to Eq.(\ref{eqn:tau_c}).

{{\bf Figure} \ref{P0_diss.fig} :}
Plot of looping probability, $P(r_0;S)$, (left) and dissociation time, $\tau_o$, (right) 
as a function of interior loop contour length $S$. 
The capture radius $r_0=5nm$.
The simulations are performed for both end-to-end ($S/L=1$) (black circle) and 
interior-to-interior ($S/L=0.5$) (red triangle). 
The parameters of the semiflexible chains are the same as in Fig.\ref{fpt_s.fig}. 
Note the loop dissociation time is much shorter than $\tau_{IC}$. 

{{\bf Figure} \ref{fpt_hetero.fig} :}
Interior looping time ($\tau_{IC}$) as a function of the reduced distance 
between sites $s_1$ and $s_2$ for the dangling ends of $s/l=0.3$ for WLC with variable persistence lengths 
($l_p^H/l_p^{(0)}=2$ in green, triangle up, $l_p^H/l_p^{(0)}=1$ in red thick line, and $l_p^H/l_p^{(0)}=0.5$ in blue, triangle down).
The physical situation corresponds to Fig.\ref{looping.fig} in which the persistence length of the 
two handles $(0,s_1)$ and $(s_2,L)$ is $l_p^H$ and the contour lengths of three segments are identical. 
For comparison the cyclization time ($\tau_c$ with $s/l=1$ and the chain persistence $l_p^{(0)}$) 
as a function of $s(=l=L/l_p^{(0)})$ is shown in black dashed line. 
The inset shows the ratios of looping times ($\tau_{IC}$) 
for the chain with variable persistence length with respect to the looping time ($\tau_{IC}^o$) of the chain with uniform persistence. 
The up triangle in green is for the ratio between $l_p^H/l_p^{(0)}=2$, $s/l=0.3$ and $l_p^H/l_p^{(0)}=1$, $s/l=0.3$ (stiff handle), 
and the dpwm triangle down in blue is for the ratio between $l_p^H/l_p^{(0)}=0.5$, $s/l=0.3$ and $l_p^H/l_p^{(0)}=1$, $s/l=0.3$ (soft handle). 

{{\bf Figure} \ref{fpt_conc.fig} :}
Plots of cyclization time ($\tau_c$) as a function of $L$ (expressed in terms of $l_p^{(0)}$) 
for various salt concentrations.
The same parameters ($l_p^{(0)}$, $D$, $\alpha$) 
with those in Fig.\ref{fpt_s.fig} are used. 
The inset shows $\tau_c$ at $L/l_p^{(0)}=3$ as a function of $c$. 
The cyclization time 
$\tau_c$ increases sharply below $c\lesssim 50mM$.

{{\bf Figure} \ref{fpt_s_combo.fig} :}
Interior looping time ($\tau_{IC}$) as a function of the reduced distance 
between site $s_1$ and $s_2$ for various size of dangling ends under two salt concentrations. 
The length of the dangling end is given by the parameter $s/l(=|s_1-s_2|/L)$.
The same parameters ($l_p^{(0)}$, $D$, $\alpha$) as 
in Fig.\ref{fpt_s.fig} are used. 
The values of the salt concentration (10mM, 500mM) are explicitly shown. 
At each value of $c$ the different curves correspond to distinct values of $s/l$. 
The values of $s/l$ range from 0.2 to 1. 
The inset shows $\tau_{IC}$, at the two values of $c$, as a function of $s/l$ for $|s_2-s_2|/l_p^{(0)}=3$ (the 
vertical dashed line). 

{{\bf Figure} \ref{interaction_pr.fig} :}
(a) End-to-end radial distribution function for a semiflexible chain 
with bending rigidity ($\kappa_b=3$, $10$, $20$, $50$, and $100$ in unit of $k_BT$) for 
various non-local interaction potentials between monomers. 
The form of the potential is $V(r)=r^{-\alpha}$. with $\alpha=1,2,4,6,12$:
Results are obtained using Brownian dynamics simulation using the energy function $E_{WLC}=E_b+E_s+\sum_{i<j}V(r_{ij})$. 
(b) The effective persistence length for various potentials $V(r)$ at different values of the bare bending rigidities.

\[\]
\vspace{3cm}
\newpage
\[\]
\begin{figure}[ht]
\includegraphics[width=5.00in]{Dfunc.eps}
\caption{\label{Dfunc.fig}}
\end{figure}
\newpage
\[\]
\begin{center}
\begin{figure}[ht]
\includegraphics[width=5.00in]{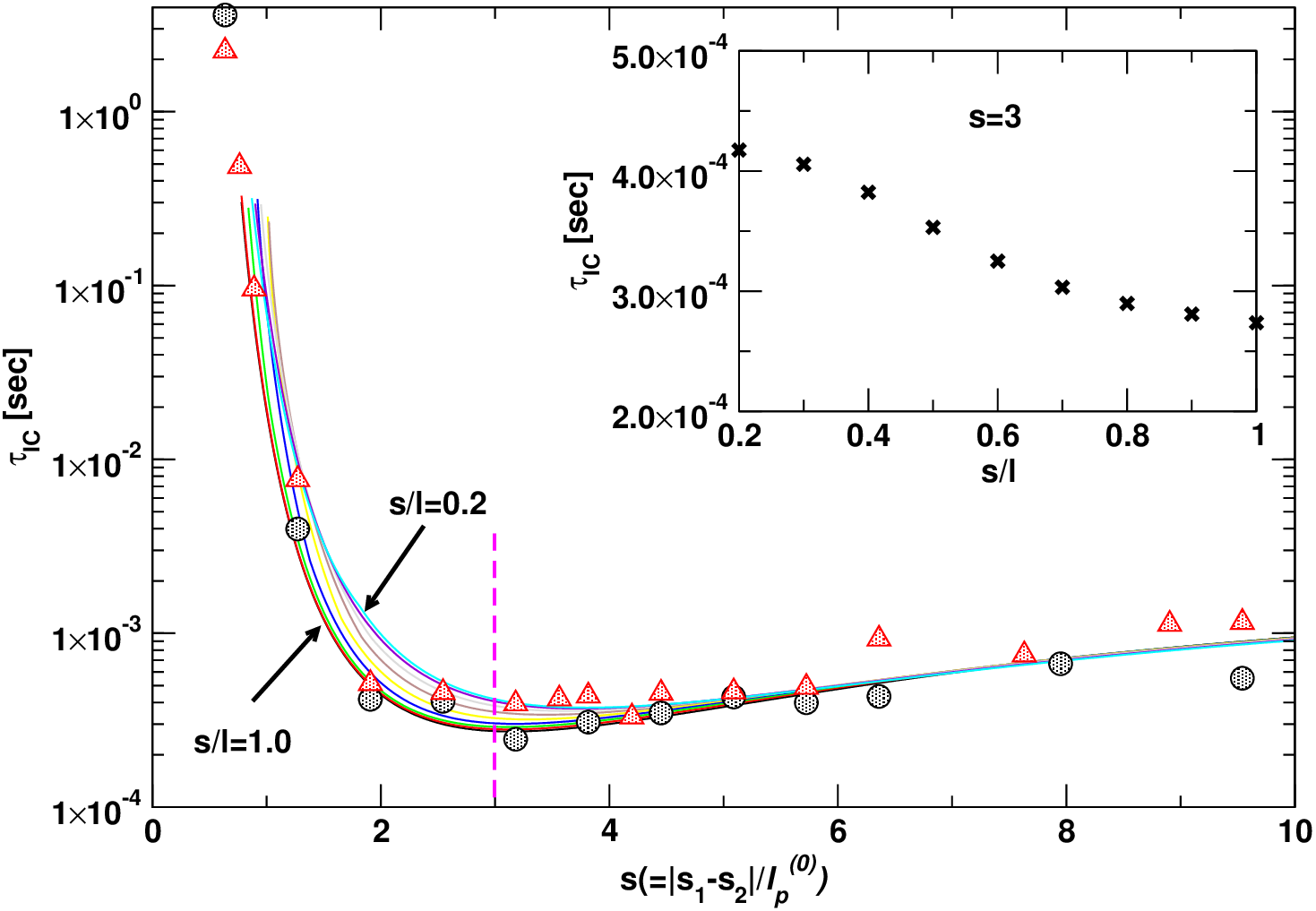}
\caption{
\label{fpt_s.fig}}
\end{figure}
\end{center}
\newpage
\[\]
\vspace{3cm}
\begin{center}
\begin{figure}[ht]
\includegraphics[width=7.00in]{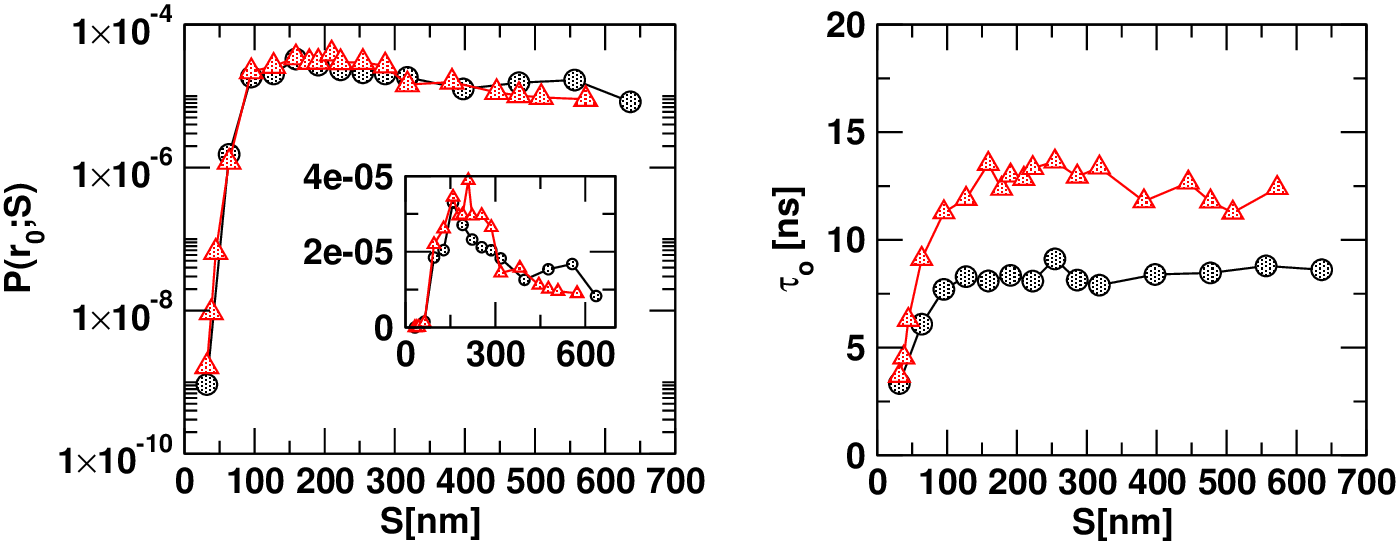}
\caption{\label{P0_diss.fig}}
\end{figure}
\end{center}

\newpage
\vspace{3cm}
\begin{center}
\begin{figure}[ht]
\includegraphics[width=5.00in]{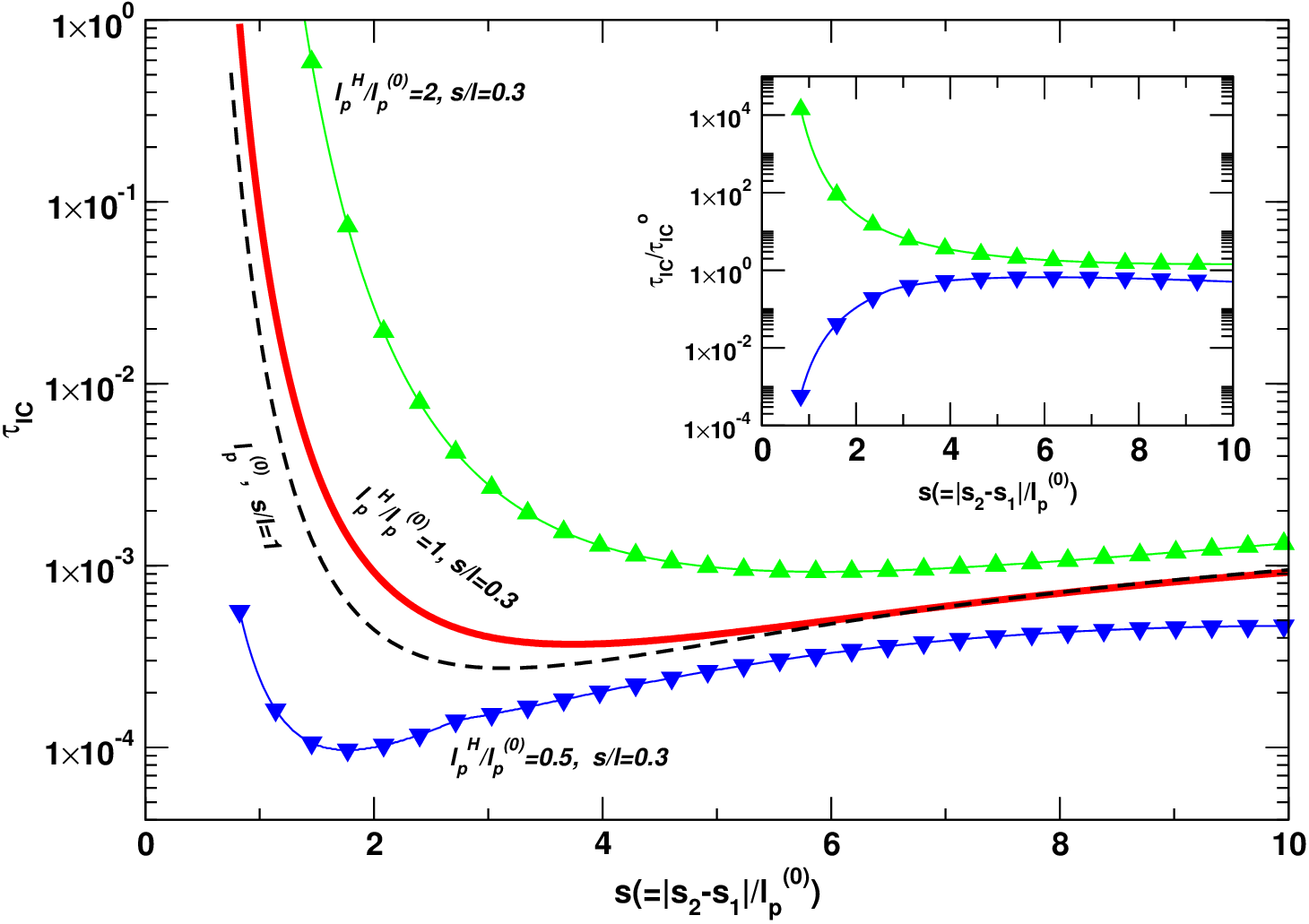}
\caption{\label{fpt_hetero.fig}}
\end{figure}
\end{center}

\newpage
\[\]
\begin{figure}[ht]
\includegraphics[width=5.00in]{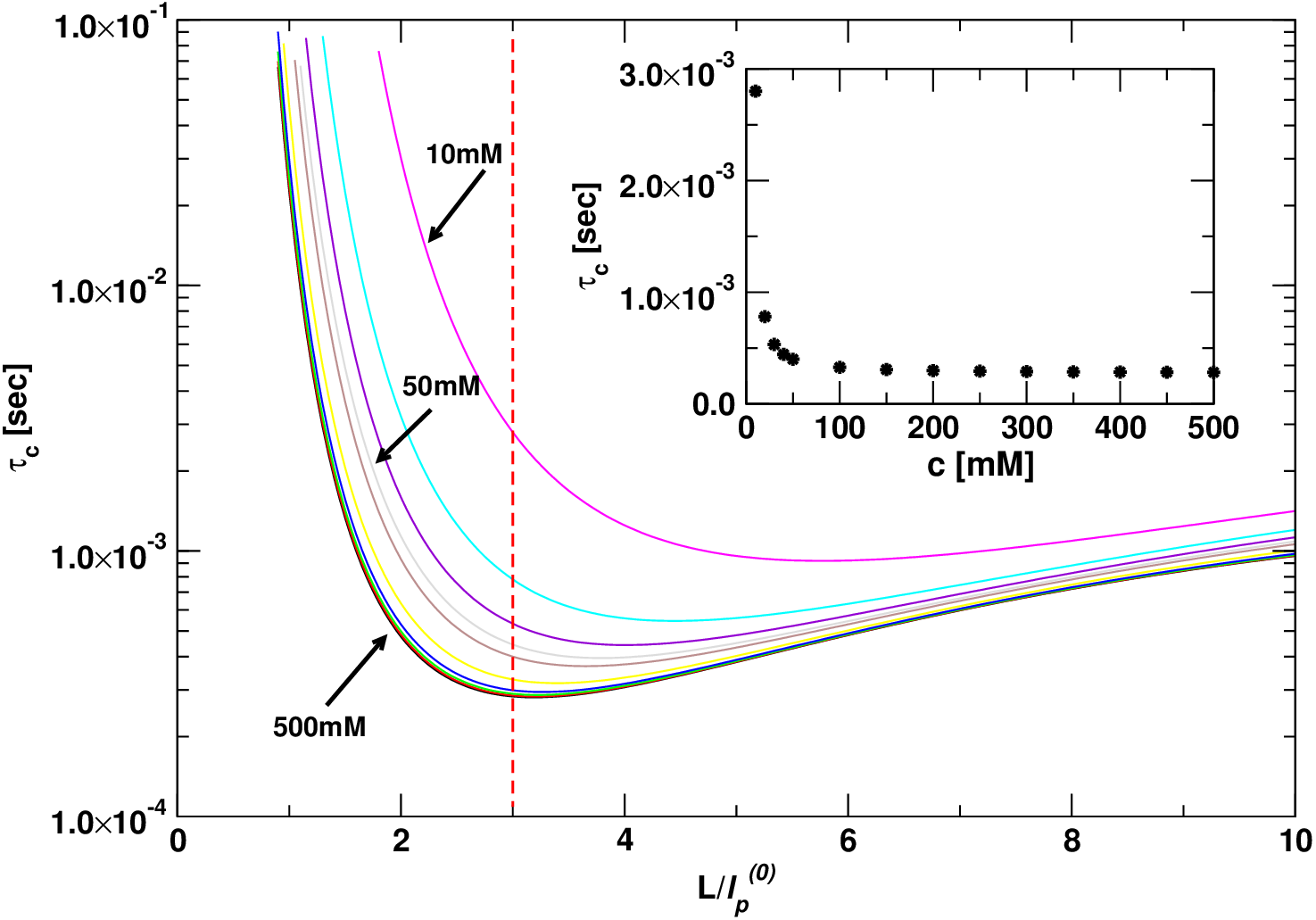}
\caption{ \label{fpt_conc.fig}}
\end{figure}

\newpage
\[\]
\begin{figure}[ht]
\includegraphics[width=5.00in]{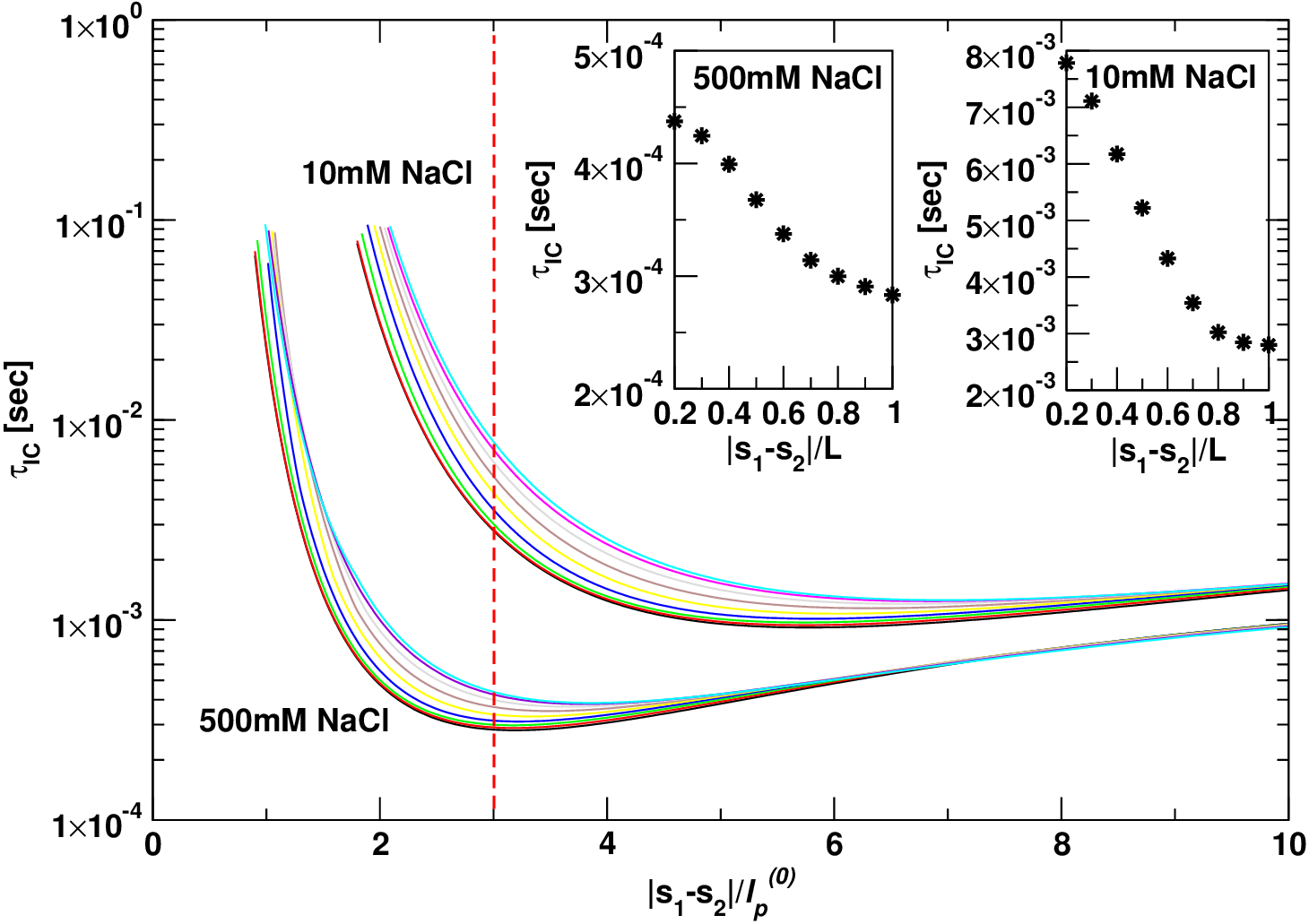}
\caption{\label{fpt_s_combo.fig}}
\end{figure}

\begin{figure}[ht]
 \mbox{
 \subfigure[]{\includegraphics[width=3.50in]{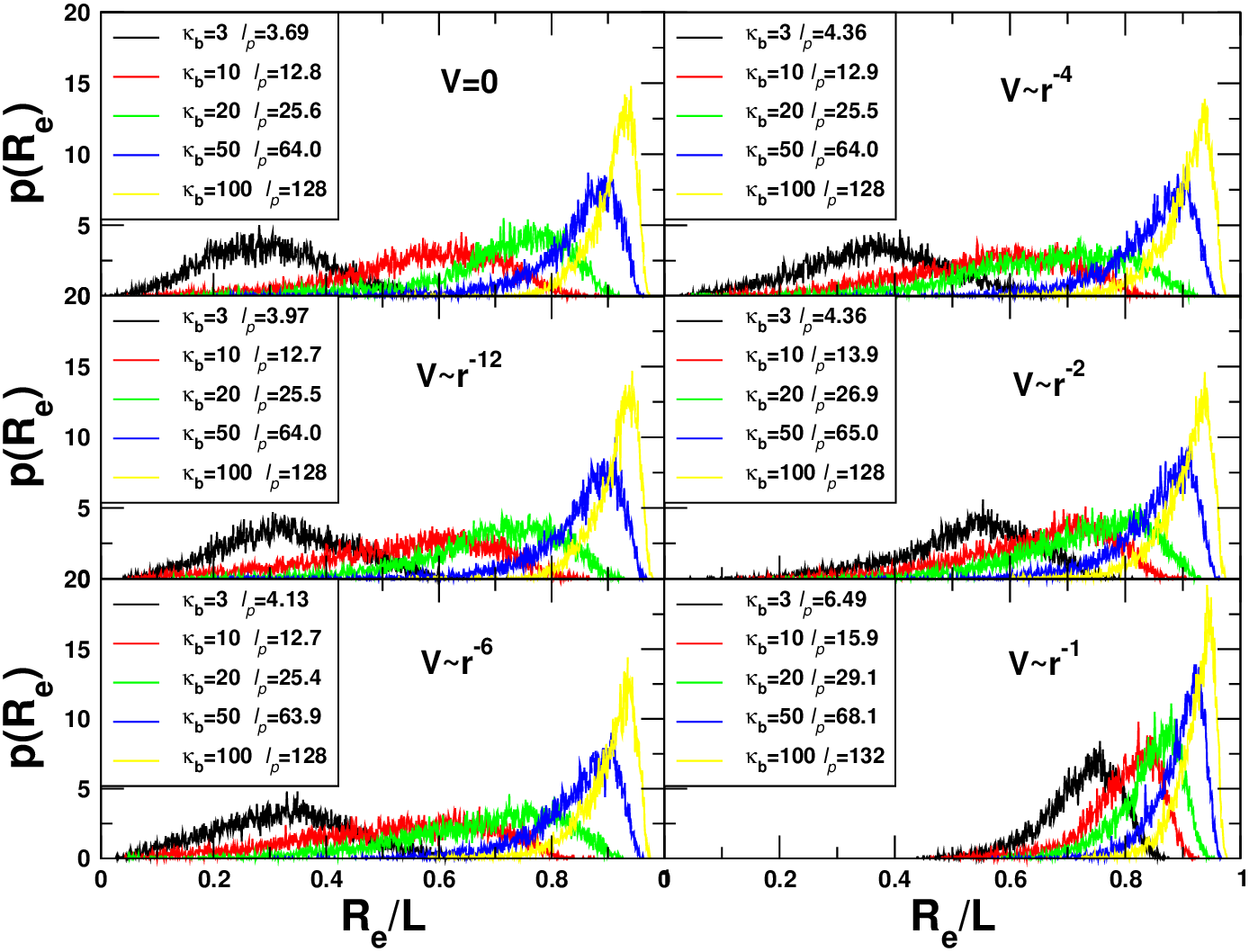}}
 }\vspace{.25in}
 \mbox{
 \subfigure[]{\includegraphics[width=3.50in]{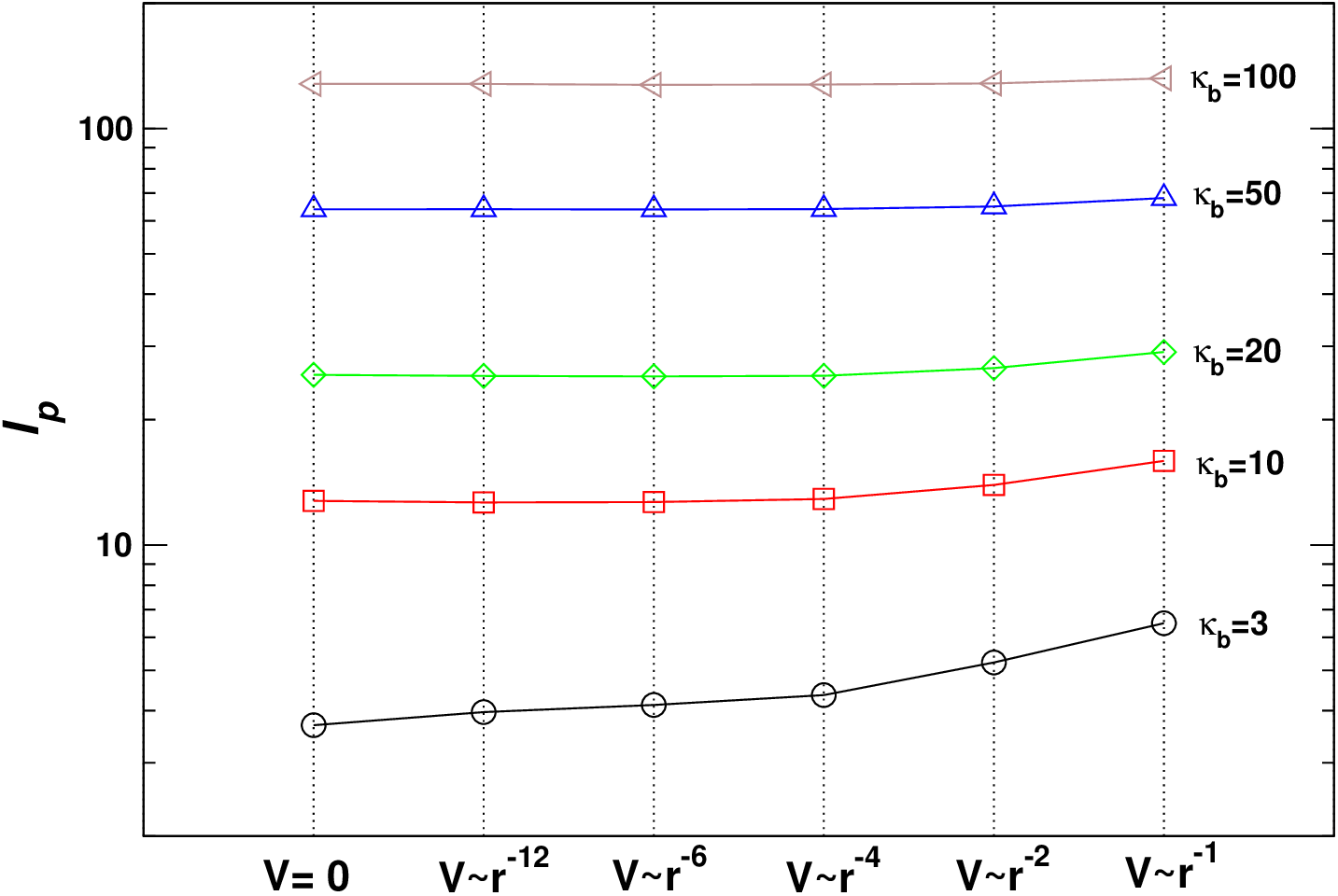}}
 }
\caption{\label{interaction_pr.fig}}
\end{figure}
\end{document}